\documentclass{pasa}%
\usepackage[utf8]{inputenc}
\usepackage{graphicx}
\def\roughly#1{\mathrel{\raise.3ex\hbox{$#1$\kern-.75em%
\lower1ex\hbox{$\sim$}}}}
\def\lsim{\roughly<}

\title[Hyperons in hot dense matter]{Hyperons in hot dense matter: what do the constraints tell us for equation of state?}

\author[Fortin et al.]{M. Fortin$^1$, M. Oertel$^2$ and C. Provid\^encia$^3$\thanks{e-mail:cp@teor.fis.uc.pt}
\affil{$^1$ N. Copernicus Astronomical Center, Polish Academy of Sciences, Bartycka 18, 00-716 Warszawa, Poland}%
\affil{$^2$LUTH, Observatoire de Paris, PSL Research University,
  CNRS, Universit\'e Paris Diderot, Sorbonne Paris Cit\'e, 5 place
  Jules Janssen, 92195 Meudon, France }
\affil{$^3$ CFisUC, Department of Physics, University of Coimbra,
  3004-516 Coimbra, Portugal}
}%

\jid{PASA}
\doi{10.1017/pas.\the\year.xxx}
\jyear{\the\year}

\usepackage{aas_macros}
\usepackage{hyperref} 
\hypersetup{colorlinks,citecolor=blue,linkcolor=blue,urlcolor=blue}

\hypersetup{draft}

\begin{document}

\begin{frontmatter}
\maketitle

\begin{abstract}
For core-collapse and neutron star merger simulations, it is important
to have at hand adequate equations of state which describe the
underlying dense and hot matter as realistically as possible.  Here,
we present two newly constructed equation of state (EoS) including the
entire baryon octet.  The two EoS are compatible with the main
constraints coming from nuclear physics, both experimental and
theoretical.  Besides, one of the EoS also describes cold
$\beta$-equilibrated neutron stars with a maximum mass of $2 M_\odot$,
in agreement with recent observations.  The predictions obtained with
the new EoS are compared with the results obtained with DD2Y. The
latter is the only presently existing EoS model containing the baryon
octet which also satisfies the above constraints within the existing
uncertainties.  The main difference between our new EoS models and
DD2Y is the harder symmetry energy of the latter.  We show that the
density dependence of the symmetry energy has a direct influence on
the amount of strangeness inside hot and dense matter and,
consequently, on thermodynamic quantities, e.g. the temperature for a
given entropy per baryon.  We expect that these differences affect the
evolution of a protoneutron star or binary neutron star mergers. We
also propose several parametrizations calibrated to $\Lambda$
hypernuclei based on the DD2 and SFHo models that satisfy the two
solar mass constraint.
\end{abstract}

\end{frontmatter}

\section{INTRODUCTION}
\label{sec:intro}
With the first detection in gravitational and electromagnetic
waves~\citep{Abbott2017,Abbott2017b}, binary neutron star mergers
promise to become outstanding sources of information for gravitational
physics, astrophysics and nuclear physics. Concerning the latter,
already for decades, neutron stars represent a formidable source to
improve our understanding of the properties of ultra-dense strongly
interacting matter. This information will be complemented by the
existing and upcoming observations of NS mergers. From the
gravitational wave (GW) signal, the masses of the two objects can be
determined. Additional information on matter properties can
potentially be obtained from the tidal deformability during late
inspiral~\citep{Faber_12,Read_13}, and, in particular, from the post
merger
oscillations~\citep{sekiguchi11,Bauswein2012,Takami2015,Bauswein2016}.
The observation of a correlated electromagnetic signal may also bring
information on matter properties. There are indications that short
gamma-ray bursts are only produced if a black hole is formed rapidly
after merger~\citep{Fryer2015,Lawrence2015}, depending thus on the
stability of the postmerger massive neutron star.  A kilonova or a
macronova event associated with the radioactive decay of produced
r-process elements is determined by matter composition and ejecta
masses~\citep{Hotokezaka13}.  In a similar way, dissipative processes
in the hot post-merger phase are controlled by matter composition and
are crucial for the evolution of the merger remnant, see
e.g. \cite{Alford2017,Fujibayashi2017}.

Whereas neutron stars are essentially formed by catalysed cold
$\beta$-equilibrated matter, temperatures as high as 50 MeV are
expected during the evolution of the post merger massive neutron star
and its envelope~\citep{sekiguchi11}. The simulation of a merger
event, where in addition $\beta$-equilibrium is not always achieved,
requires, therefore, an equation of state (EoS) of warm stellar matter
in a wide range of temperatures, electron fractions and
densities. Also, directly after its formation in a core-collapse
supernova (CCSN), the newly born proto-neutron star evolves from matter with
a quite large proton content to neutron rich matter, emitting large
amounts of neutrinos and reaching temperatures of up to 100 MeV
\citep{Burrows1986,Keil1995,Pons1999}. Thus, for core-collapse
simulations such an EoS\footnote{We will call these EoS models
  ``general purpose EoS'' in the following text.}  covering
temperatures of $0 \lsim T \lsim 100$ MeV, baryon number densities
$10^{-11} \mathrm{fm}^{-3} \lsim n_B \lsim 10 \, \mathrm{fm}^{-3}$
as
well as electron fractions $0 \lsim Y_e = n_e/n_B \lsim 0.6$ is also
essential.

The EoS by \cite{Lattimer:1991nc} ( ``LS'') and that by
\cite{Shen:1998by} (``STOS'') are the two most widely used general
purpose models in simulations. Much effort has been devoted in the
last years to improve these two classical models, see
\cite{OertelRMP16} for a recent review. First, several EoS models have
been proposed which improve the treatment of non-homogeneous matter
containing nuclear clusters at low densities and temperatures, for
instance
by~\cite{Hempel09,Raduta:2010ym,hempel12,steiner13,Gulminelli:2015csa}. Second,
at high densities and temperatures non-nucleonic degrees of freedom
such as hyperons, mesons or quarks have been included into the EoS
models, see
e.g.~\cite{ishizuka_08,Nakazato08,sagert_09,Shen:2011qu,Oertel:2012qd,Banik:2014qja}.

However, additional degrees of freedom in the EoS lower the maximum
neutron star mass and it was only recently that the first EoS model
(``DD2Y'') was proposed~\citep{Marques2017} containing the whole
baryonic octet and being able to describe a cold neutron star with a
mass of 2$ M_\odot$ in agreement with
observations~\citep{Demorest2010,Fonseca2016,Antoniadis2013}. The
underlying nuclear model (DD2, \cite{Typel:2009sy}) satisfies the
presently accepted constraints on nuclear matter at saturation density
and below coming from nuclear experiments and theoretical
calculations. Having in mind that it is important that simulations of
supernovas or binary neutron star merging should include realistic
EoS, and that accepted properties of nuclear matter are still defined
within an interval of uncertainties, we propose in the present work
two other EoS models containing the entire baryonic octet. These EoS
have as underlying nuclear model SFHo~\citep{steiner13}, a model which also
satisfies the accepted constraints on nuclear matter. The main
difference to DD2 is the smaller symmetry energy, leading among
others, to a considerably lower prediction for the radius of a
fiducial 1.4 $M_\odot$  star:
11.9 km instead of 13.3 km
obtained with DD2. Here, we
will analyse how this property will influence the amount of hyperons
inside hot and dense matter and the impact on thermodynamic
properties.

Considering cold $\beta-$equilibrated NS, in general hyperons appear
only in stars with a mass above roughly 1.4 $M_\odot$, and then
usually the overall fractions remain small if the 2$M_\odot$
maximum mass constraint is imposed, see for instance
\cite{Fortin2016}. It is, therefore, expected that they will not play
a major role in determining global NS quantities
such as the radii, moments of inertia and tidal
deformabilities. We emphasize, however, that the study of neutron
stars, their formation process in core-collapse supernovae and their
mergers is not restricted to the cold $\beta$-equilibrated equation of
state. Matter is heated up a lot in core-collapse and in the
post-merger phase. Besides, the evolution of these processes require
the knowledge of transport properties such as the specific heat, viscosity
or neutrino emissivities that do depend on the constitutents
of matter. Hyperons may, as well,  play an important role in
NS cooling.
 It has been shown in \cite{Fortin2016} that several EoS that
satisfy experimental and theoretical constraints up to saturation density do not allow for
the fast cooling direct Urca process. In these cases, direct URCA processes
only operate when hyperons appear, and, therefore, hyperons
may have a strong influence in cooling, see also the recent
  study in \cite{raduta2017}. 
Hyperons may also strongly affect the proto-neutron star cooling and stability and lead in particular to a delayed stellar mass black-hole formation, potentially observable via the associated neutrino signal
(see e.g. \cite{Keil1995}, \cite{Prakash1997}, \cite{Pons2001},
\cite{Peres_13}).  In neutron star mergers too, hyperons may leave a
clear imprint, see for instance \cite{sekiguchi11a}. Note,
  however, that the employed EoS does not fulfill the $2M_\odot$
constraint and hence that more work is necessary. Considering
future neutron star merger detections, it is important to test all
factors that might influence the final result, and hyperons is one of
these factors.

The paper is organised as follows: { in Sec.
  \ref{sec:eos} the model for describing dense matter are presented,
  in Sec.  \ref{sec:constraints} the properties of the two new
  hyperonic EoS based on SFHo are discussed and compared with model
  DD2Y. In the last section we summarise our results. Technical
  details on how the matching between clusterized and uniform
  matter is done, as well as technical issues concerning the EoS
  Tables in the COMPOSE data base are given in the appendices.  }

\section{MODEL FOR THE EQUATION OF STATE}
\label{sec:eos}
Most available general purpose EoS models including the entire baryon
octet and covering at the same time a sufficiently large range in
baryon number density, $n_B$, temperature $T$ and hadronic charge
fraction, $Y_Q = n_Q/n_B = Y_e$~\footnote{$n_Q$ represents the total
  hadronic charge density.}  in order to be applicable in CCSN or
binary mergers, are either not compatible with constraints from
nuclear physics and/or a neutron star maximum mass of $2 M_\odot$, see
the discussion in~\cite{OertelRMP16}.  We will compare here
two different EoSs, considering both all hyperons and being well
compatible with the main present constraints: the DD2Y
EoS~\citep{Marques2017} and a new EoS based on the nuclear SFHo
EoS~\citep{steiner13}, see below for details.

\subsection{Description of inhomogeneous matter}
The main aim of the present paper is to discuss the appearance of
hyperons in high density/high temperature homogeneous matter and its
impact on the EoS. In order to obtain a unified EoS over the entire
needed range in temperature, baryon number density and charge
fraction, the present EoS models are combined with a description of
inhomogeneous clustered matter at densities below roughly saturation
density and low temperatures based on the respective nuclear
interaction, using the statistical model by Hempel \&
Schaffner-Bielich~\citep{Hempel09,hempel12,steiner13}. For a more detailed
discussion of the issues related to clustered nuclear matter in
stellar environments, see
e.g.~\cite{Raduta:2010ym,Gulminelli:2015csa,
  Hempel09,Typel:2009sy, Sumiyoshi08a, Buyukcizmeci_14, Heckel09}.

\subsection{Homogeneous matter}
\label{sec:homogeneous}
We will treat homogeneous matter within two different phenomenological
relativistic mean field (RMF) models. Baryonic interactions are
modelled by the exchange of ``meson'' fields. The term ``meson'' refers
thereby to the quantum numbers of the different interaction channels.
The literature on those models is large and many different
parameterizations exist (see e.g.~\cite{Dutra2014}).

We will use one model with density dependent
couplings and one model with non-linear couplings of the meson
fields. The Lagrangian density is written in the following
form
\begin{eqnarray} {\mathcal L} &=& \sum_{j \in \mathcal{B}}  \bar
  \psi_j \left( i\gamma_\mu \partial^\mu - m_j + g_{\sigma j} \sigma 
\right. \\ & -& \left.
    \,g_{\omega j} \gamma_\mu \omega^\mu -  \,g_{\phi j} \gamma_\mu
    \phi^\mu - \, g_{\rho j} \gamma_\mu \vec{\rho}^\mu \cdot
    \vec{I}_j\right) \psi_j \nonumber \\ &+& \frac{1}{2}
  (\partial_\mu \sigma \partial^\mu \sigma - m_\sigma^2 \sigma^2)
 - \frac{g_2}{3} \sigma^3 - \frac{g_3}{4} \sigma^4 \nonumber \\ & 
 -& \frac{1}{4} W^\dagger_{\mu\nu} W^{\mu\nu} - \frac{1}{4}
  P^\dagger_{\mu\nu} P^{\mu\nu} - \frac{1}{4} \vec{R}^\dagger_{\mu\nu}
  \cdot \vec{R}^{\mu\nu} \nonumber \\ & + &\frac{1}{2} m^2_\omega
  \omega_\mu \omega^\mu + \frac{1}{2} m^2_\phi
  \phi_\mu \phi^\mu + \frac{1}{2} m^2_\rho \vec{\rho}_\mu \cdot
  \vec{\rho}^\mu\nonumber \\
& +& \frac{c_3}{4} (\omega_\mu\omega^\mu)^2 + \frac{c_4}{4}
     (\vec{\rho}_\mu \vec{\rho}^\mu)^2 + A(\sigma,\omega^\mu
     \omega_\mu) \vec{\rho}^{\mu}\cdot\vec{\rho}_\mu  ~,\nonumber 
\end{eqnarray}
where $\psi_j$ denotes the field of baryon $j$, and $W_{\mu\nu},
P_{\mu\nu}, \vec{R}_{\mu\nu}$ are the field tensors of the vector
mesons, $\omega$ (isoscalar), $\phi$ (isoscalar), and $\rho$
(isovector), of the form
\begin{eqnarray}
V^{\mu\nu} = \partial^\mu  V^\nu - \partial^\nu V^\mu~.
\end{eqnarray}
$\sigma$ is a  scalar-isoscalar meson field. 
For the baryon masses $m_j$ we have taken the following values:
$m_n = 939.565346$, $m_p = 938.272013$, and  $m_\Lambda = 1115.683,$ $m_\Sigma =
1190$, $m_{\Xi^-} = 1321.68$ ,$m_{\Xi^0} = 1314.83$ MeV ($m_\Lambda = 1116.0$, $m_{\Sigma^{+,0,-}} =
1189.0, 1193.0, 1197.0$,  $m_{\Xi^-} = 1321.0$, $m_{\Xi^0} = 1315.0$ MeV) for DD2Y (SFHoY).

The couplings of meson $M$ to baryon $j$ are conveniently written in the following form within models with density dependence,
\begin{equation}
  g_{Mj}(n_B) = g_{Mj}(n_0) h_M(x)~,\quad x = n_B/n_0~.
\end{equation}
The density $n_0$ is thereby a normalization constant, usually
taken to be the saturation density $n_0 = n_{\mathit{sat}}$ of
symmetric nuclear matter. 
Here, we will consider the DD2
parameterization~\citep{Typel:2009sy}, where the functions $h_M$ assume the following form for the isoscalar couplings~\citep{Typel:2009sy},
\begin{equation}
h_M(x) = a_M \frac{1 + b_M ( x + d_M)^2}{1 + c_M (x + d_M)^2}
\end{equation}
and
\begin{equation}
h_M(x) = \exp[-a_M (x-1)] ~
\end{equation}
 for the isovector ones. See~\cite{Typel:2009sy} for the values of
the parameters $a_M, b_M, c_M,$ and $d_M$. The coupling constants of
the nonlinear terms, $g_2,g_3,c_3,c_4$ and the function $A$
are absent in models with density-dependent couplings.

The function 
\begin{equation}
A(\sigma,\omega_\mu \omega^\mu) = \sum_{i=1}^6 a_i \sigma^i + \sum_{j = 1}^3 b_j (\omega_\mu \omega^\mu)^j
\end{equation}
in addition to the nonzero couplings $g_2,g_3,c_3,$ and $c_4$ has been
introduced in~\cite{Steiner2005} ($A=g_\rho^2 f$ with $f$ defined in
eq. (15) of this reference) to be able to vary easily the
symmetry energy in nonlinear models. Here, we employ the SFHo
parameterization~\citep{steiner13}.

Both models will be used in mean field approximation, where the meson fields are replaced by their respective expectation values in uniform matter:
\begin{eqnarray}
m_\sigma^2 \bar\sigma &=& \sum_{j \in B} g_{\sigma j} n_j^s - g_2 \bar\sigma^2 - g_3 \bar\sigma^3 + \frac{\partial A}{\partial \bar\sigma} \bar\rho^2
\\
m_\omega^2 \bar\omega &=& \sum_{j \in B} g_{\omega j} n_j - c_3 \bar\omega^3 - \frac{\partial A}{\partial \bar\omega} \bar\rho^2 \\
m_\phi^2 \bar\phi &=& \sum_{j \in B} g_{\phi j} n_j\\
m_\rho^2 \bar\rho &=& \sum_{j \in B} g_{\rho i} t_{3 j} n_j - c_4 \bar\rho^3 - 2 \, A\, \bar\rho~,
\end{eqnarray}
where $\bar\rho=\langle\rho_3^0\rangle$,
$\bar\omega=\langle\omega^0\rangle$, $\bar \phi=\langle\phi^0\rangle$,
and $t_{3 j}$ represents the third component of isospin of baryon $j$
with the convention that $t_{3 p} = 1/2$. The scalar density of baryon
$j$ is given by
\begin{equation}
  n^s_j = \langle \bar \psi_j \psi_j \rangle = \frac{1}{\pi^2} \int
  k^2 \frac{M^*_j} {\epsilon_j(k)} \{f[\epsilon_j(k)]
  +\bar{f}[ \epsilon_j(k)] \} dk~,
\end{equation}
and the number density by
\begin{equation}
n_j = \langle \bar \psi_j\gamma^0 \psi_j \rangle = \frac{1}{\pi^2} \int
k^2 \{f[\epsilon_j(k)] - \bar{f}[\epsilon_j(k)]\}dk ~.
\end{equation}
$f$ and $\bar{f}$ represent here the occupation numbers of the
respective particle and antiparticle states with $\epsilon_j(k) =
\sqrt{k^2 + M^{*2}_j}$, and effective chemical potentials
$\mu^*_j$. They reduce to a step function at zero
temperature. The effective baryon mass $M^*_j$ depends on the scalar
mean fields as
\begin{equation}
M^*_j = M_j - g_{\sigma j} \bar\sigma~,
\end{equation}
and
the effective chemical potentials
are related to the chemical potentials via
\begin{equation}
\mu_j^* = \mu_j - g_{\omega j} \bar\omega - g_{\rho j} \,t_{3 j}
\bar\rho - g_{\phi j} \bar \phi - \Sigma_0^R~.
\label{mui}
\end{equation}
The rearrangement term $\Sigma_0^R$ is present in models with
density-dependent couplings to ensure thermodynamic consistency. It is
given by
\begin{eqnarray}
  \Sigma_0^R &=& \sum_{j \in B} \left( \frac{\partial g_{\omega j}}{\partial n_j}
                 \bar\omega n_j + t_{3 j} \frac{\partial g_{\rho j}}{\partial n_j}
                 \bar\rho n_j +\frac{\partial g_{\phi j}}{\partial n_j}
                 \bar\phi n_j \right. \nonumber \\
             && \left. -\frac{\partial g_{\sigma j}}{\partial n_j}
                \bar\sigma n_j^s  \right)~.
\end{eqnarray}

In contrast to the nuclear interaction which can be well constrained
up to saturation density by information on nuclear properties, the
information from hypernuclei is scarce and does not allow to fix the
parameters of the model. In many recent
works~\citep{Weissenborn11c,Banik:2014qja,Miyatsu:2013hea}, the
isoscalar vector meson-baryon coupling constants are hence related following
a symmetry inspired procedure such that the couplings of hyperons to
isoscalar vector mesons are expressed in terms of $g_{\omega N}$ and a
few additional parameters, see e.g.~\cite{Schaffner96}.  In general,
an underlying $SU(6)$-symmetry and ideal $\omega$-$\phi$-mixing is
assumed, completely fixing the hyperonic couplings in terms of
$g_{\omega N}$.  Extending the above procedure to the isovector sector
would lead to contradictions with the observed nuclear symmetry
energy. $g_{\rho N}$ is therefore left as a free parameter and the
remaining hyperonic isovector couplings are fixed by isospin symmetry.

This procedure has been adopted for the DD2Y model~\citep{Marques2017},
but for the SFHo model with hyperons additional repulsion is needed
such that the EoS for cold neutron star matter remains compatible with
a maximum mass of $2 M_\odot$ as required by observations, see
Section~\ref{sec:constraints}. We therefore rescale the $\omega$ and
$\phi$-meson hyperon couplings as follows: $ g_{M\Lambda} = 1.5
g_{M\Lambda}(SU(6)), g_{M\Sigma} = 1.5 g_{M\Sigma}(SU(6)), g_{M\Xi} =
1.875 g_{M\Xi}(SU(6))$. This EoS model will be called ``SFHoY''. In
addition we will discuss the model with $SU(6)$ couplings, called
``SFHoY$^*$''.

Comparing properties of single-$\Lambda$-hyperons with data on
{ single $\Lambda$} hypernuclei then allows to determine the remaining
scalar coupling~\citep{vandalen_14,Fortin2017}. Less data are available
for $\Xi$ and $\Sigma$. An alternative, although less precise, way is
to to use the values of hyperonic single-particle mean field
potentials to constrain the scalar coupling constants.  The potential
for particle $j$ in $k$-particle matter is given by
\begin{equation}
U_j^{(k)}(n_k) = M^*_j - M_j + \mu_j - \mu^*_j~.
\label{Ujk}
\end{equation}
We use here standard values at nuclear matter saturation density,
$n_{\mathit{sat}}$~\citep{Weissenborn11c,Fortin2017},
$U_\Lambda^{(N)}(n_{\mathit{sat}})= -30$ MeV,
$U_{\Xi}^{(N)}(n_{\mathit{sat}}) = -18$ MeV (DD2Y) and $-14$ MeV
(SFHoY), and $U_\Sigma^{(N)} (n_{\mathit{sat}}) = + 30$ MeV. In
section~\ref{sec:hypernuclei} we will show that, in view of the
uncertainties on the data, the obtained couplings are compatible with
hypernuclear data.

Table~\ref{tab:couplings} summarizes the values of the
meson hyperon couplings in both models obtained from the
above described procedure.
\begin{table*}
\begin{tabular}{l|cccccccccccc}
\hline
Model & $R_{\sigma \Lambda}$ & $R_{\omega \Lambda}$ & $R_{\phi \Lambda}$ & $R_{\rho \Lambda}$ 
&$R_{\sigma\Sigma}$
  &$R_{\omega \Sigma}$ &   $R_{\phi \Sigma}$ &   $R_{\rho \Sigma}$ &   
$R_{\sigma \Xi}$ &$R_{\omega \Xi}$ &$R_{\phi \Xi}$ &$R_{\rho \Xi}$ 
\\
DD2Y & 0.62 & 2/3 &-0.47 &0 & 0.48 &2/3 &-0.47 &2 & 0.32 & 0.33 &-0.94 &1 \\
SFHoY & 0.85 & 1 &-0.71 &0 & 0.58 &1 &-0.71 &2  & 0.51 &0.62 &-1.77 & 1\\
SFHoY$^*$ &0.61  & 2/3 &-0.47 &0 &0.35  &2/3 &-0.47 &2  & 0.30 &0.33
                                   &-0.94 & 1\\
\hline
\end{tabular}
\caption{\label{tab:couplings} Coupling constants of the mesons to
  different hyperons within the three models presented here, normalized
  to the respective meson nucleon coupling, i.e. $R_{M j} = g_{M j}/
  g_{M N}$, except for the $\phi$-meson. Here the $g_{\omega N}$ is
  used for normalisation. }
\end{table*}

\section{EQUATION OF STATE PROPERTIES}
\label{sec:constraints}
\subsection{Summary of constraints}
Any model for the EoS has to be confronted with various constraints:
\begin{itemize}
\item 
The recent observation of two massive neutron stars, indicating the
maximum mass of a cold, non- or slowly-rotating (therefore spherically
symmetric) neutron star should be above $2 M_\odot$ gives a very
robust constraint on the interactions at supra-saturation densities.
\item Laboratory experiments on finite nuclei can constrain the EoS up
  to roughly saturation density. The main sources of information are
  nuclear mass measurements, neutron skin data, nuclear resonances,
  dipole polarizability of nuclei, nuclear decays, and heavy ion
  collisions. Experimental data can - within a model used for the
  analysis- be correlated with nuclear matter properties, which are in
  general chosen as the coefficients of a Taylor expansion of the
  energy per baryon of isospin symmetric nuclear matter around
  saturation.  Values with a reasonable precision can be obtained for
  the saturation density ($n_{\rm sat}$), binding energy ($E_B$),
  incompressibility ($K$), symmetry energy ($E_{\rm sym}$) and its
  slope ($L$). The extracted values depend of course on the model used
  for the analysis. In some cases, it has recently been shown that this
  model dependence can be reduced if values at $n_B = 0.1
  \mathrm{fm}^{-3}$ are given instead of saturation density~\citep{Khan:2013mga}.
\item Much effort has recently been devoted to theoretical ab-initio
  calculations of pure neutron matter in order to constrain the
  equation of state and roughly up to saturation density good
  agreement between the different approaches has been achieved. Since
  compact stars contain neutron rich matter, this information is very
  interesting and completes the constraints on symmetric matter.
\end{itemize}
A summary and discussion of the most important available constraints
can be found e.g. in \cite{OertelRMP16}.

\begin{figure}
\includegraphics[width=1.0\columnwidth]{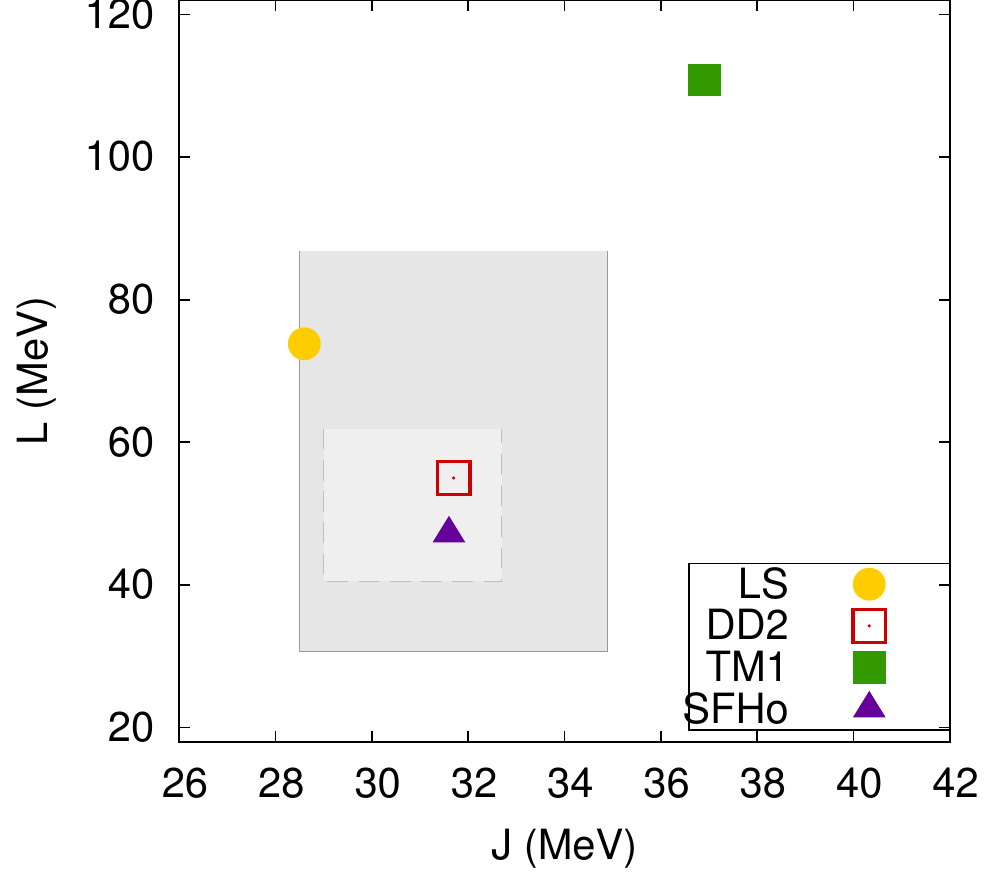}
\caption{(color online) Values of $E_{\rm sym}$ and $L$ in different nuclear
  interaction models. The two gray rectangles correspond to the range
  for $E_{\rm sym}$ and $L$ derived in~\cite{Lattimer:2012xj} (light Gray)
  and~\cite{OertelRMP16} (dark gray) from nuclear experiments and
  some neutron star observations.
  \label{fig:jl}}
\end{figure}
The two parameterizations chosen in this paper as basis for the EoS
models, DD2 and SFHo, both agree reasonably well with most of the
established constraints, see Table~\ref{tab:nmatter} for the values of
different nuclear matter properties. For comparison we show the values
for two other interactions, that of the Lattimer and Swesty EoS
(LS)~\citep{Lattimer:1991nc} and that for the TM1
parameterization~\citep{Sugahara_94}, too. These two interactions have
been employed in other recently developed general purpose EoS,
including non-nucleonic degrees of freedom,
e.g.~\cite{ishizuka_08,Shen:2011qu,Oertel:2012qd}.
\begin{table}
\begin{center}
\begin{tabular}{l|ccccc}
\hline
 Model& $n_{\rm sat}$  & $E_B$ & $K$ & $E_{\rm sym}$& $L$  \\
&$(\mathrm{fm}^{-3})$ & (MeV) & (MeV) &(MeV) & (MeV) \\ \hline \hline
DD2         &0.149   & 16.0 & 243& 31.7 &55.0 \\
SFHo& 0.158&16.2 & 245&31.6&47.1\\
LS220& 0.155&16.0 & 220&28.6&73.8\\
TM1& 0.145&16.3 & 281&36.9&110.8\\
\hline

\end{tabular}
\caption{Nuclear matter properties of the two nuclear interaction models used within the different EoS models. For comparison, the corresponding values of the interactions of the two standard EoS models, LS~\citep{Lattimer:1991nc} and TM1~\citep{Sugahara_94} are also given. }
\label{tab:nmatter}
\end{center}
\end{table}
\begin{figure}
\includegraphics[width=1.0\columnwidth]{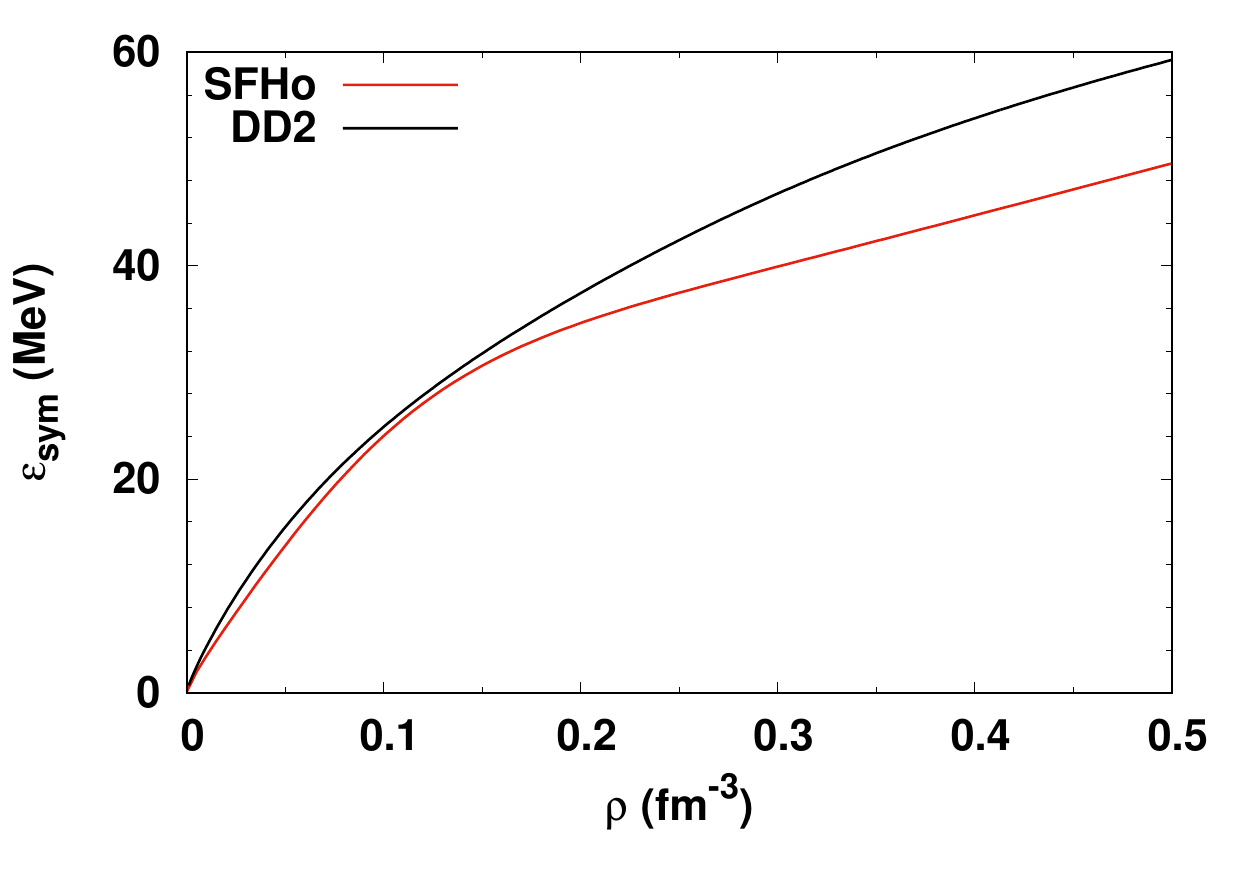}
\caption{(color online) Symmetry energy as function of baryon number density within the two parameterisations employed here: SFHo and DD2. 
\label{fig:esym}}
\end{figure}
Saturation density, binding energy and incompressibility of the two
interactions used here lie within standard ranges~\citep{OertelRMP16}:
$n_{\rm sat}\simeq 0.15-0.16$ fm$^{-3}$, $E_B\simeq 16$ MeV and
$K=248\pm 8$ MeV \citep{Piek04} and $K=(240\pm 20)$ MeV
\citep{Shlomo06}. Note that in \cite{Stone2014} higher values of the incompressibility in the range 250-315 MeV are favored. The last word is thus not said, but given the uncertainties, DD2 and SFHo still lie in within a reasonable range. The compatibility of $E_{\rm sym}$ and $L$ with
ranges derived in~\cite{Lattimer:2012xj} (light gray rectangle) and
in~\cite{OertelRMP16} (dark gray rectangle), respectively, are shown
in Fig.~\ref{fig:jl}. In contrast to LS and TM1, the values of the two
interactions employed here, DD2 and SFHo, are situated well within the
rectangles.

In Fig.~\ref{fig:esym} we display the density dependence of the two
parameterisations, DD2 and SFHo, we have used as a basis to construct
our EoS models. It is evident that DD2 produces a larger symmetry
energy throughout the entire relevant density range.  Neither the DD2
nor the SFHo allow for the nucleonic direct Urca process. This was
already noted for DD2 in \cite{Fortin2016}. Moreover, the authors of
\cite{Fortin2016} identify the value $L=70$ MeV as a limiting value
below which the nucleonic direct Urca process would occur only in
stars with a mass above 1.5$M_\odot$, and possibly simply not occur,
depending on the model. The latter is the case for the considered RMF
models that satisfy the constraints coming from microscopic
calculations for neutron matter and the constraints on the saturation
properties of nuclear matter. An exception is NL3$\omega\rho$ which,
however, only allows nucleonic direct Urca in stars with a mass above
$2.5 M_\odot$.  It is, therefore, not surprising that SFHo also shows
a similar behavior, since it satisfies the same constraints.
With the opening of the hyperonic degrees of freedom, the hyperonic
direct Urca becomes possible. For DD2 this is true for stars with $M>
1.52 M_\odot$ with the onset of the $\Lambda$. For SFHoY (SFHoY*) the
onset of the hyperon $\Xi^-$ ($\Lambda$) defines the onset of the
hyperonic direct Urca and occurs for stars with $M>1.56M_\odot$
($M>1.26M_\odot$). For SFHoY the $\Lambda$ only sets in if
$M>1.70M_\odot$. It is expected that hyperon superfluidity will
suppress neutrino emissivities due to hyperonic direct Urca
processes. However, the superfluid hyperon gap is strongly dependent
on the hyperon properties in neutron star matter and on the $YY$
interaction \citep{wang2010}. In particular, it has been shown the
small value of $\Delta B_{\Lambda\Lambda}$ obtained from
$^6_{\Lambda\Lambda}$He \citep{nagara} leads to a very small gap in
dense neutron matter \citep{tanigawa2003} and in dense neutron star
matter \citep{wang2010} or even suppresses completely the $\Lambda$
superfluidity. Following \cite{wang2010}, $\Lambda$ superfluidity
exists in massive neutron stars only for strong $\Lambda\Lambda$
interactions. For the other hyperons, uncertainties are even larger,
see also the discussion in \cite{raduta2017}.
Therefore, at the present stage it is not possible to conclude which could be
the role of hyperon superfluidity on the neutrino emissivity.

\begin{figure*}
\includegraphics[width=1.0\linewidth]{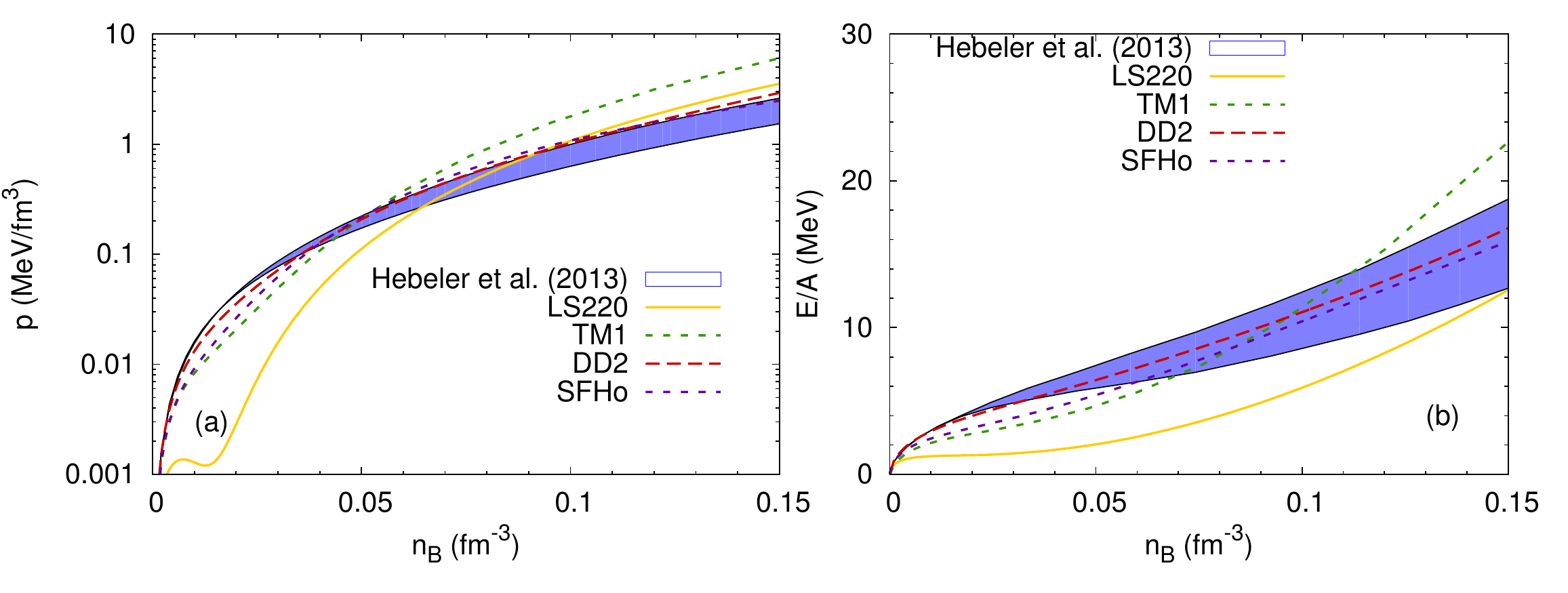}
\caption{(color online) Pressure (left panel) and energy per baryon
  (right panel) of pure neutron matter as functions of baryon number
  density within different nuclear interaction models compared with
  the ab-initio calculations of~\cite{Hebeler:2013nza}, indicated
  by the blue band.
\label{fig:nm}}
\end{figure*}
As mentioned above, ab-initio calculations of pure neutron matter can
serve as a constraint on the EoS, too. Hence, in Fig.~\ref{fig:nm} we
show pressure and energy per baryon for pure neutron matter, comparing
results from the different nuclear interactions with the ab initio
calculations from~\cite{Hebeler:2013nza}, including an estimate of the
corresponding uncertainties. None of the displayed models is in perfect
agreement with the theoretical calculations, however, DD2 and SFHo
show much better agreement than the standard LS and TM1 models. At
densities below roughly $n_B = 0.1 \mathrm{fm}^{-3}$, where the
deviations of our models with the theoretical predictions are largest,
one could in addition argue that the EoS of stellar matter is anyway
strongly influenced by the treatment of nuclear clusters (nuclear
masses, surface effects, thermal excitations, \dots) and the
interaction model is not very important, see e.g. the discussion
in~\cite{OertelRMP16}.

\begin{figure}{!h}
\includegraphics[width=1.0\columnwidth]{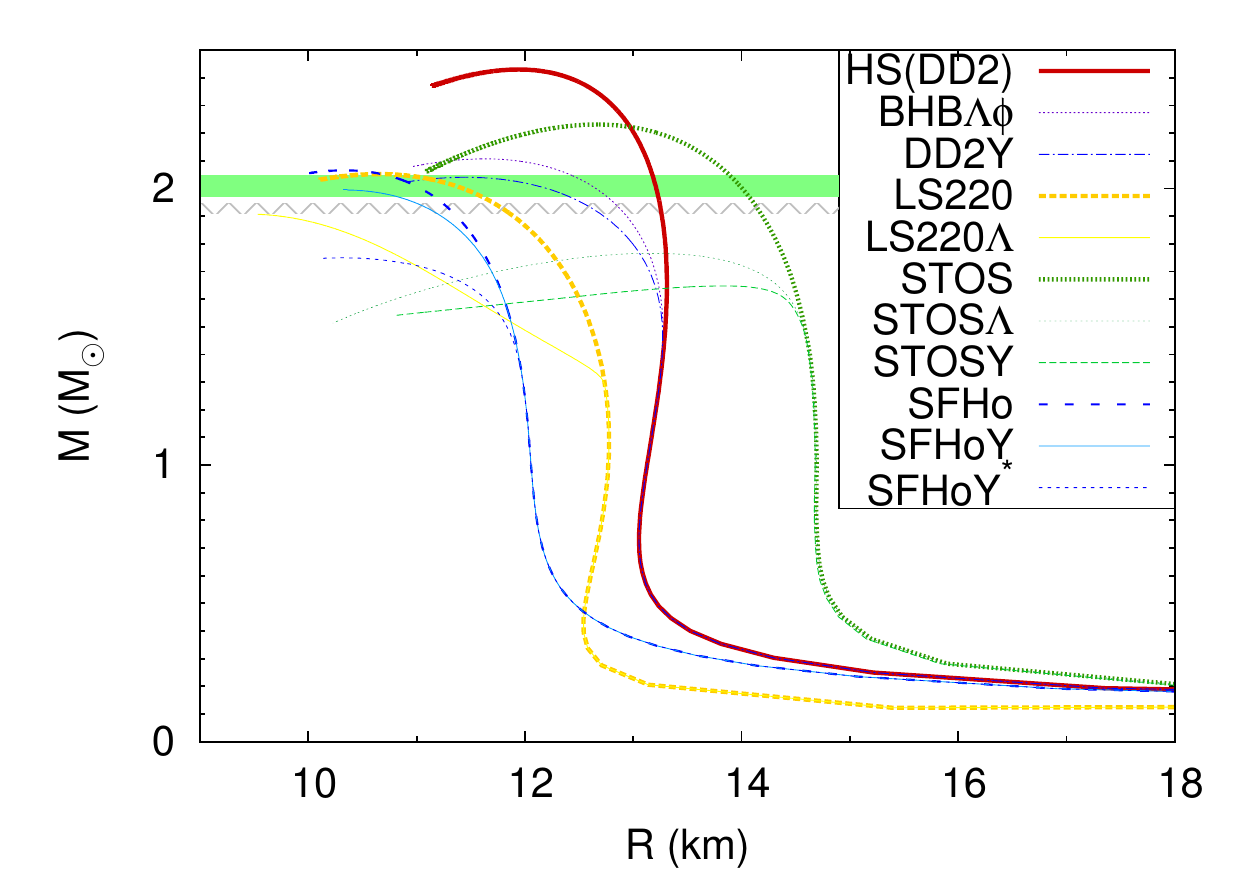}
\caption{(color online) Gravitational mass versus circumferential
  equatorial radius for cold spherically
  symmetric neutron stars within different EoS models. The two
  horizontal bars indicate the two recent precise NS mass
  determinations, PSR J1614-2230~\citep{Demorest2010,Fonseca2016}
  (hatched gray) and PSR J0348+0432~\citep{Antoniadis2013} (green).
\label{fig:mreos}}
\end{figure}

In Fig.~\ref{fig:mreos} we display the mass-radius relation of
cold\footnote{For convenience we have chosen a temperature of $T =
  0.1$ keV for producing this figure. In the following discussion of
  our results we always refer to this temperature upon speaking about
  ``cold'' stars.}  spherically symmetric neutron stars within
different general purpose EoS models.  In addition to the models
containing the entire baryon octet discussed here,
DD2Y~\citep{Marques2017}, SFHoY, and SFHoY$^*$, we show the purely
nuclear LS EoS~\citep{Lattimer:1991nc}, its extension with
$\Lambda$-hyperons (``LS220$\Lambda$'')~\citep{Peres_13}, the nuclear
EoS by Shen et al. (``STOS'') employing the TM1
interaction~\citep{Shen:1998by}, its extension with $\Lambda$-hyperons
(``STOS$\Lambda$'') \citep{Shen:2011qu} and all hyperons
(``STOSY'')~\citep{ishizuka_08}, as well as one model including
$\Lambda$-hyperons within DD2 from~\cite{Banik:2014qja}
(``BHB$\Lambda\phi$''). It is evident that among the EoS containing
hyperons, apart from BHB$\Lambda\phi$ which contains only
$\Lambda$-hyperons, only the two EoS DD2Y and SFHoY are compatible
with the $2 M_\odot$-constraint.  A summary of cold neutron star
properties for the different EoSs is given in
Table~\ref{tab:nsresultsT0}.
\begin{figure*}[h!]
\begin{tabular}{cc}
	\includegraphics[width=1.\columnwidth]{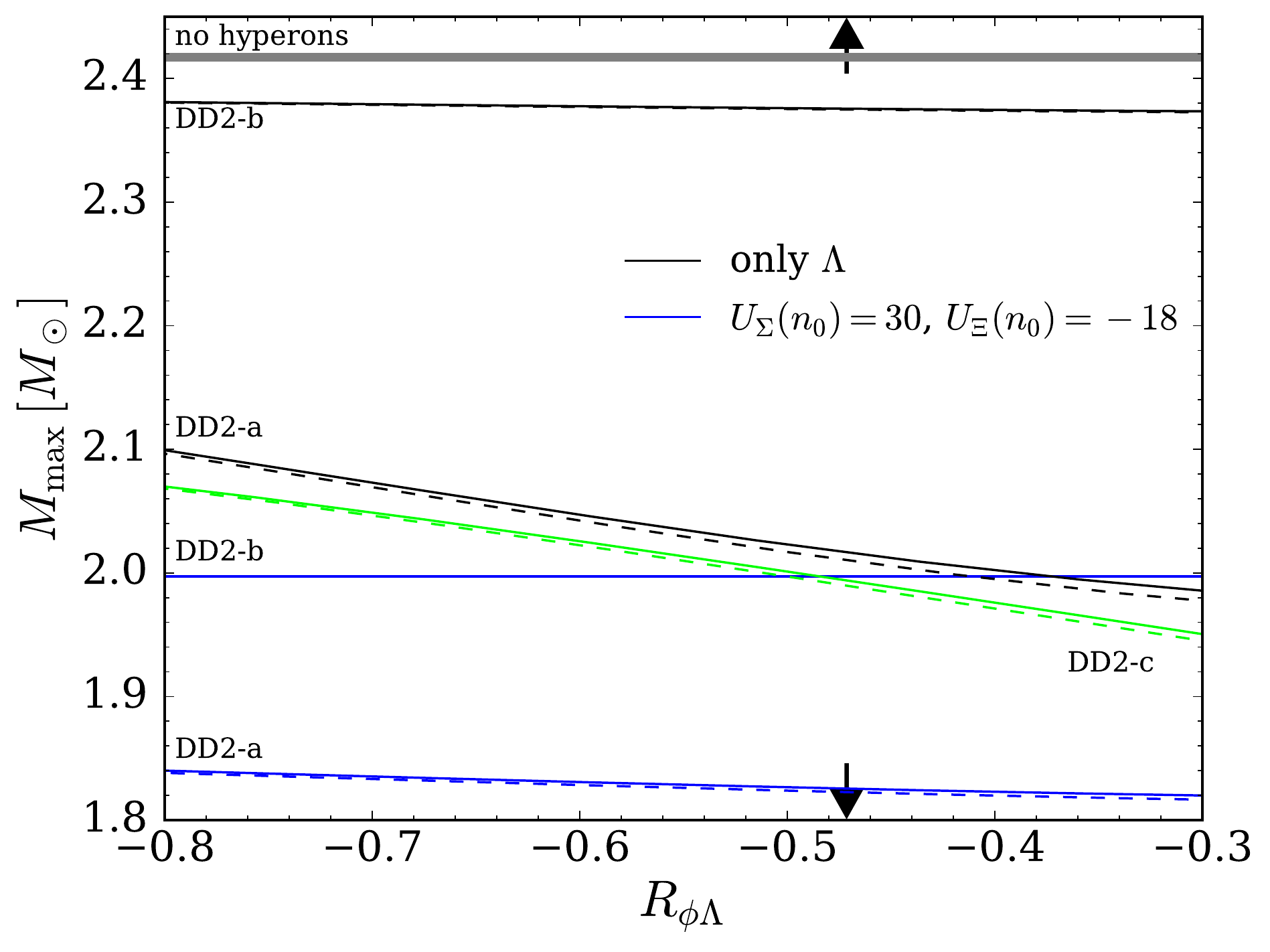}&
	\includegraphics[width=1.\columnwidth]{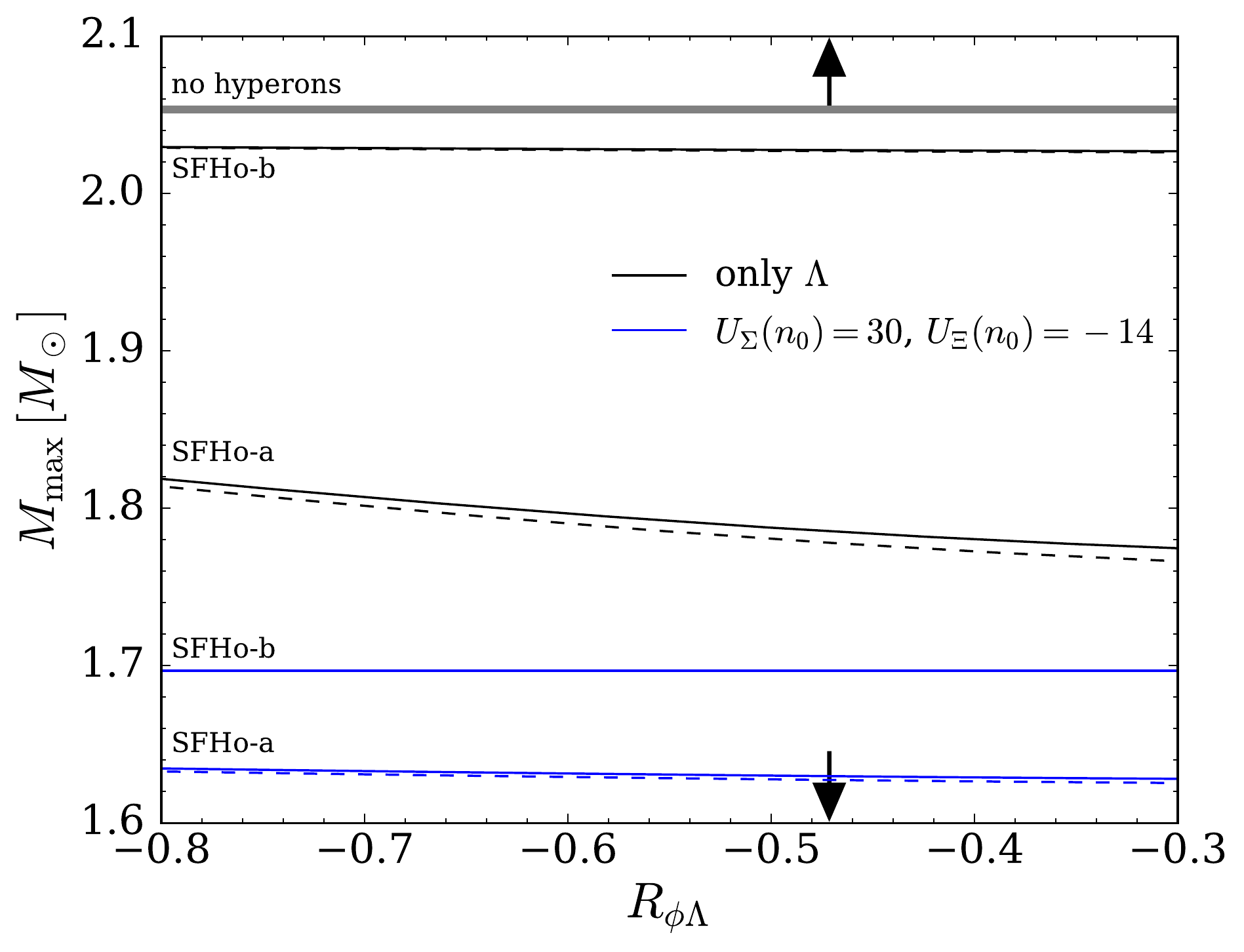}\\
	\includegraphics[width=1.\columnwidth]{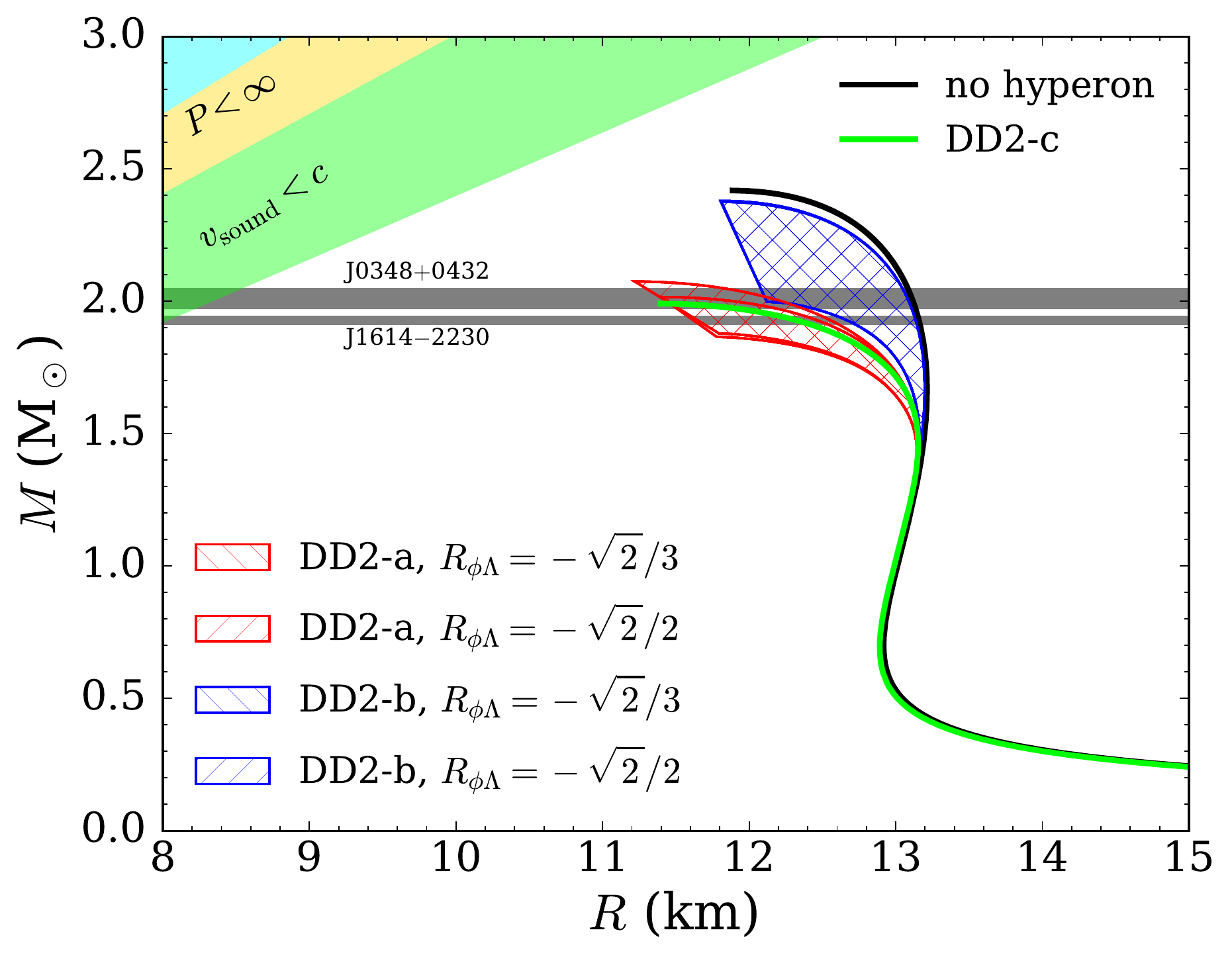}&
	\includegraphics[width=1.\columnwidth]{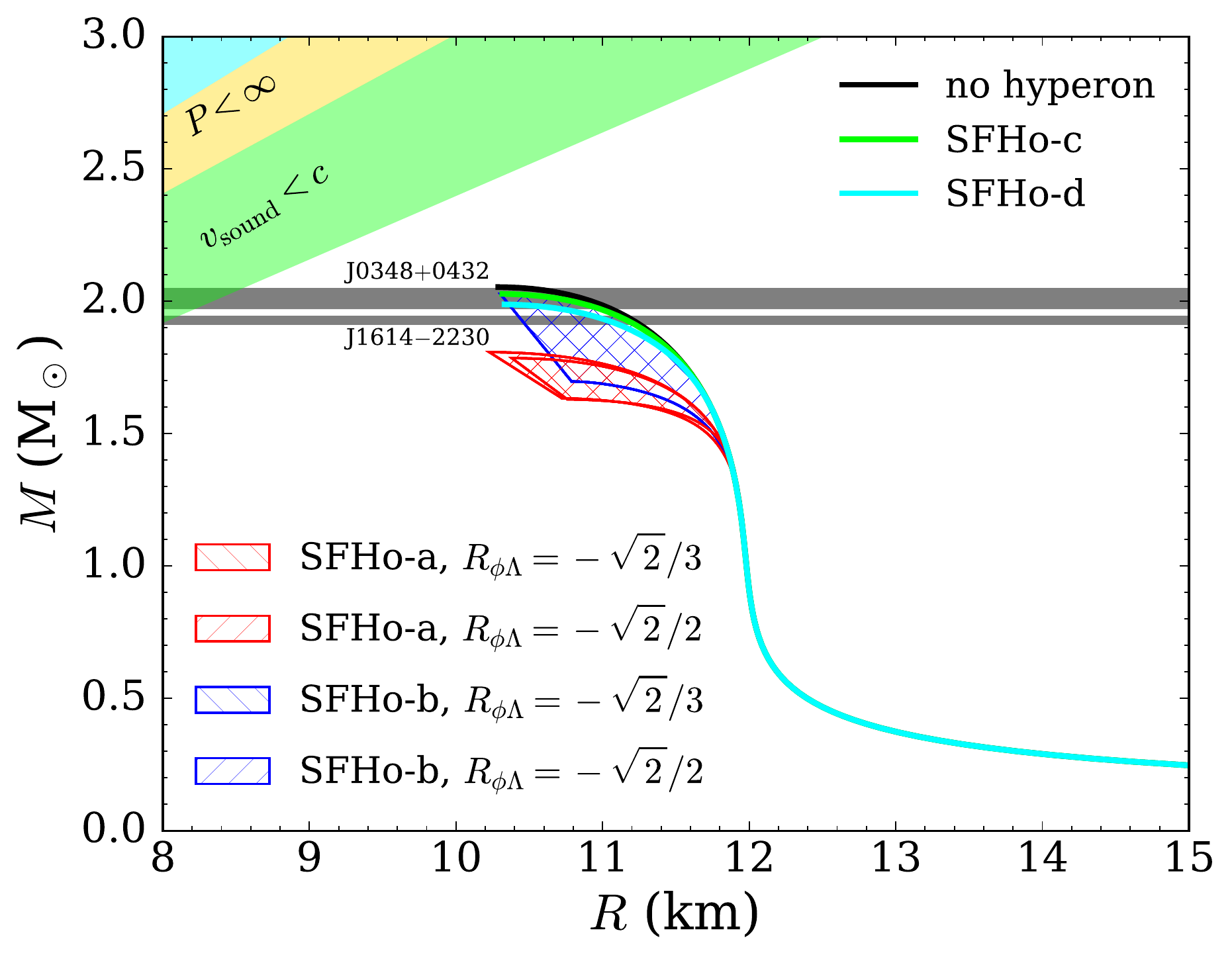}
\end{tabular}
    \caption{ DD2-x (left) and SFHo-x (right) parametrizations,  top panels: neutron star maximum
      mass $M_{\rm max}$ as a function of $R_{\phi\Lambda}$ for
      various hyperonic models. The values $R_{\sigma\Lambda}$,
      $R_{\phi\Lambda}$ and $R_{\sigma^{\ast}\Lambda}$ are adjusted to
      reproduce the binding energies of single $\Lambda$-hypernuclei
      and of $_{\Lambda \Lambda}^6$He with $\Delta B_{\Lambda
        \Lambda}=0.50$ MeV (solid lines) and 0.84 MeV (dashed
      lines). The arrows indicate the SU(6) value of
      $R_{\phi\Lambda}$ { and the gray line the maximum mass for a purely nucleonic model}; 
      bottom panels: MR curves for the
      parametrizations obtained { for the models -a ie. taking $R_{\Lambda\omega}=2/3$ (red
      region) and the models -b that is with $R_{\Lambda\omega}=1$} (blue region) for the two
      different values of $R_{\Lambda\phi}$ indicated in the Table
      \ref{tab:double}. In all cases the upper limit is defined
    including only $\Lambda$s   and the bottom line including the
    complete baryonic octet with the couplings chosen as explained in
    the text. The black line is for pure nucleonic stars 
      and  the green
    line is for the parametrization of  the set  DD2$-c$ (left)  
 or  SFHo$-c$ (right), see the text for details. In the bottom right panel the cyan
line identified as SFHo$-d$ was obtained with the calibrated
$\sigma-\Lambda$ parameters for  $R_{\Lambda\omega}=1$ and the
couplings to the $\Sigma$ and $\Xi$ as in the SFHoY model}. 
   \label{fig:DD2hypernuclei}
\end{figure*}
\begin{table}
\begin{center}
\begin{tabular}{l|ccccc}
\hline
 Model& $M_g^{\mathit{max}}$  & $M_B^{\mathit{max}}$ & $R_{1.4}$ & $f_S$& $n_B^{(c)}$  \\
&$[M_\odot]$ & [$M_\odot$]& [km]& & [fm$^{-3}$] \\ \hline \hline
HS(DD2)         &2.43   & 2.90 & 13.3 & - &0.84 \\
BHB$\Lambda\phi$  &2.11 &2.47 &13.3&0.05& 0.96\\
DD2Y & 2.04 &2.36 & 13.3 &0.04 &1.00\\
SFHo& 2.07 &2.44 & 11.9 & - &1.15\\
SFHoY& 1.99&2.36 & 11.9&0.02&1.18\\
SFHoY$^*$&1.75&2.02&11.9&0.05&1.25\\
\hline
\end{tabular}
\caption{Properties of cold spherically symmetric neutron stars in
  neutrinoless $\beta$-equilibrium: Maximum gravitational and baryonic
  masses, respectively, the total strangeness fraction, $f_S$,
  representing the integral of the strangeness fraction $Y_S/3$ over
  the whole star, defined as in~\cite{Weissenborn11c}, and the central
  baryon number density. The latter two quantities are given for the
  maximum mass configuration. In addition, the radius at a fiducial mass of $M_g =
  1.4 M_\odot$ is listed.  For comparison with the hyperonic EoS
  DD2Y, SFHoY and SFHoY$^*$, the values for the purely nucleonic EoS
  models HS(DD2)~\citep{Fischer_13} and SFHo~\citep{steiner13}, as
  well as the the $BHB\Lambda\phi$ EoS including $\Lambda$-hyperons
  based on HS(DD2) from~\cite{Banik:2014qja} are also given.  }
\label{tab:nsresultsT0}
\end{center}
\end{table}

\begin{figure*}
\includegraphics[width=0.9\linewidth]{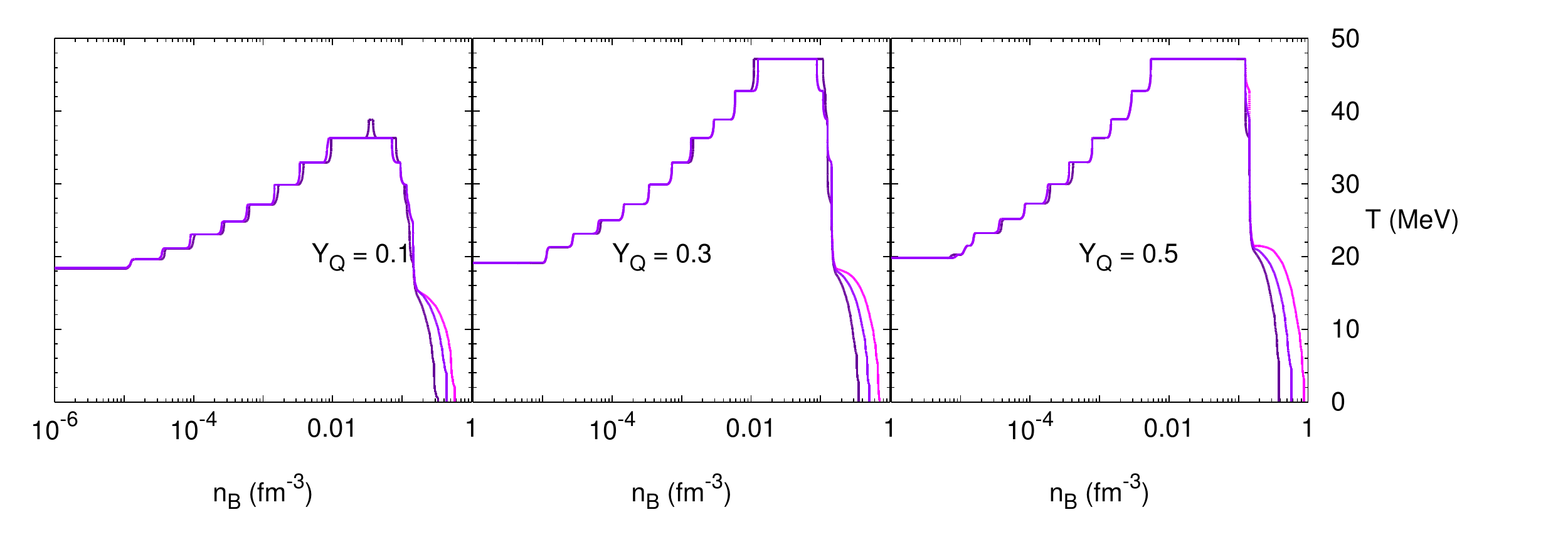}
\caption{(color online) The lines delimit the regions in temperature
  and baryon number density for which the overall hyperon fraction
  exceeds $10^{-4}$, which are situated above the lines. Different
  charge fractions are shown as indicated within the panels. The lines
  correspond to DD2Y, SFHoY$^*$ and SFHoY, respectively, appearing in
  that order at low temperatures and high densities.
  \label{fig:contour}
}
\end{figure*}
\begin{figure*}
\includegraphics[width=1.\linewidth]{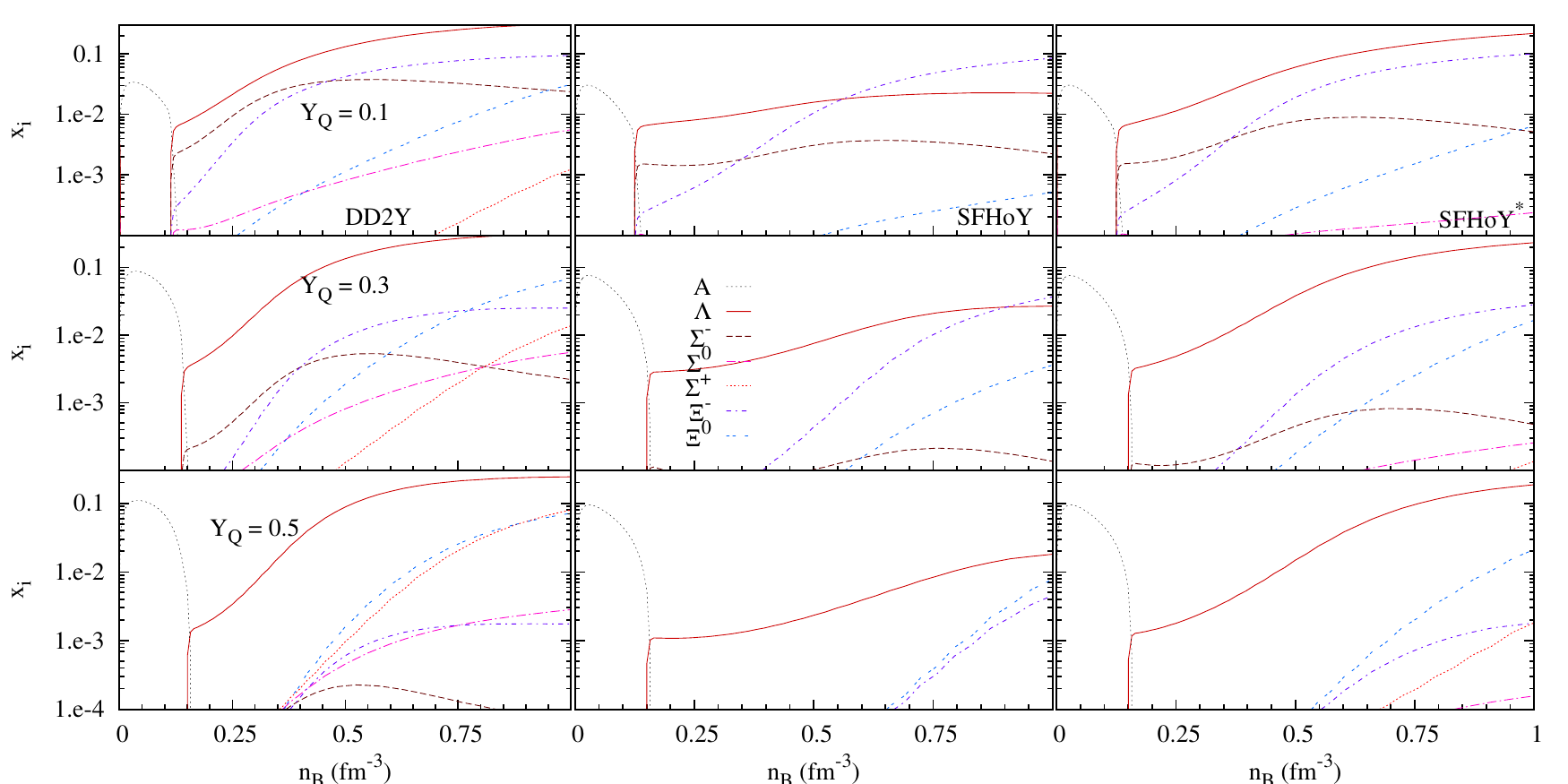}
\caption{(color online) 
  \label{fig:xifuncnb} Particle fractions versus baryonic density  for
  DD2Y (left), SFHoY (middle), and SFHoY$^*$ (right) for
  $T=30$ MeV and different charge fractions.The label
  ``A''indicates the sum over all different nuclei.
}
\end{figure*}

\subsection{Properties of $\Lambda$-hypernuclei}
\label{sec:hypernuclei}
In this section, we study to which extent the hyperonic couplings of
DD2Y and SFHoY reproduce experimental data of $\Lambda$-hypernuclei. To that
end, we follow the approach of \cite{Fortin2017}, see this reference
for more details on the calculations.
To calculate the binding energies of single and double
$\Lambda$-hypernuclei we have solved Dirac equations for the nucleons
and the $\Lambda$ using the method described in~\cite{Avancini07}.  A
tensor term is included as in~\cite{Shen06} in order to obtain a weak
$\Lambda$-nuclear spin-orbit interaction. This term has no effect in
homogeneous matter.  As mentioned in the previous section, we assume
the same density dependence for hyperon- and nucleon-meson couplings. 
\begin{table*}[ht]
\center
\begin{tabular}{cccccccccccc}
\hline
Model& &  &&  & \multicolumn{3}{c}{$\Delta B_{\Lambda \Lambda}=0.50$} &$\phantom{m}$&
                                                                      \multicolumn{3}{c}{$\Delta B_{\Lambda \Lambda}=0.84$}\\
\cline{6-8}\cline{10-12}\\
 &$R_{\omega\Lambda}$ & $R_{\sigma\Lambda}$  & $U_\Lambda^N(n_{\mathit{sat}})$ & $R_{\phi\Lambda}$ &$R_{\sigma^\ast\Lambda}$ &  $U_\Lambda^\Lambda(n_{\mathit{sat}})$ & $U_\Lambda^\Lambda(n_{\mathit{sat}}/5)$ &&$R_{\sigma^{\ast}\Lambda}$& $U_\Lambda^\Lambda(n_{\mathit{sat}})$& $U_\Lambda^\Lambda(n_{\mathit{sat}}/5)$ \\ 
\hline
&&&\\
DD2$-a$&$2/3$ & 0.623 & -31&  $-\sqrt{2}/3$ & 0.545 &-13.40 & -4.06&& 0.564 &-15.79 &-4.79\\ 
DD2$-a$&  $2/3$ & 0.623 & -31      & $-\sqrt{2}/2$ & 0.840 &-12.78 & -4.10&& 0.852 &-15.06 &-5.33\\ 
DD2$-b$& $1$ & 0.900 & -37 & $-\sqrt{2}/3$ & 0.576 &-9.63 & -4.38&& 0.600 &-12.68 &-4.81\\ 
 DD2$-b$& $1$ & 0.900 & -37        & $-\sqrt{2}/2$ & 0.860 &-6.61 & -4.36& &0.876 & -9.55&-5.33\\
SFHo$-a$&$2/3$ & 0.609 & -29&  $-\sqrt{2}/3$ & 0.461&-14.17 &-6.52&&0.501& -16.86&-7.07\\ 
SFHo$-a$&  $2/3$ & 0.609 & -29      & $-\sqrt{2}/2$ &0.751 &  -13.69& -6.56& &0.777&-16.47&-7.13\\  
SFHo$-b$& $1$ & 0.870 & -34 & $-\sqrt{2}/3$ &0.485  &-9.21&-10.58&&0.526&-12.08&-11.18\\ 
 SFHo$-b$& $1$ & 0.870 & -34        & $-\sqrt{2}/2$ & 0.767 &-8.53&-10.65&&0.793& -11.33&-11.23\\ 
\hline 
\end{tabular} 

\caption{Calibration to $\Lambda$-hypernuclei and
  $_{\Lambda\Lambda}^6$He for models with the $SU(6)$-value,
  $R_{\omega\Lambda}=2/3$ ($a$), and $R_{\omega\Lambda}=1$ ($b$).  For
  given a $R_{\omega\Lambda}$, values of $R_{\sigma\Lambda}$ are
  calibrated to reproduce the binding energies $B_\Lambda$ of single
  hypernuclei in the s- and p-shells. The value of the
  $\Lambda$-potential in symmetric baryonic matter at saturation is
  given for reference.  For given $R_{\phi\Lambda}$,
  $R_{\sigma^{\ast}\Lambda}$ are calibrated to reproduce the upper and
  lower values of the bond energy of $_{\Lambda \Lambda}^6$He. For
  reference the $\Lambda$-potential in pure $\Lambda$-matter at
  saturation and at $n_{\mathit{sat}}/5$ are also given. All energies are given in
  MeV.
\label{tab:double}}
\end{table*}

With the values of the coupling constants given in
Table~\ref{tab:couplings}, the experimental binding energies
$B_{\Lambda}$ of hypernuclei in s- and p-shells as given in Table IV
of~\cite{Gal16} are well reproduced, for both models, DD2Y and
SFHoY. The bond energy of $_{\Lambda \Lambda}^6$He, the only known
double-$\Lambda$ hypernucleus, is, on the contrary, not well
reproduced by both parameterisations. The reason is that the
$\Lambda\Lambda$ interaction is too repulsive at low densities.

 In
\cite{Marques2017} a second parameterisation, DD2Y$\sigma^*$, has been
proposed, including the hidden strangeness meson $\sigma^*$, coupling
to hyperons and rendering the hyperon-hyperon ($YY$) interaction more
attractive at low densities. Within DD2Y$\sigma^*$, the bond energy
comes out fine. But, the neutron star maximum mass, with a value of
only 1.87 $M_\odot$, does not fulfil the constraints from
observations.  
At the present stage, we will keep the DD2Y parameterisation since
developing a new hyperonic model based on the  DD2 model is beyond the scope of the present paper. But,
as indicated below, following the directions given in
\cite{Fortin2017} for other models, parameterisations can be found in
agreement with both, the $_{\Lambda \Lambda}^6$He bond energy and a
neutron star mass of $2 M_\odot$ as well as the single-$\Lambda$
hypernuclear data.

To that end, the $SU(6)$ constraint on the isoscalar vector couplings
has to be relaxed. This can be seen from
Fig.~\ref{fig:DD2hypernuclei}. For the top panels of that  figure, the ratio
$R_{\omega\Lambda}$ has been fixed, and then the ratio
$R_{\sigma\Lambda}=g_{\sigma\Lambda}/g_{\sigma N}$ has been fitted to
the experimental binding energies $B_\Lambda$ of hypernuclei in the s-
and p-shells and the coupling of the $\Lambda$ to $\sigma^*$ to the
bond energy of $_{\Lambda\Lambda}^6$He, see \cite{Fortin2017} for
details. The obtained neutron star maximum mass is then shown as a
function of the value of $R_{\phi\Lambda}$ for several different
cases. The lines labeled DD2$-a$ (SFHo$-a$) thereby indicate the $SU(6)$ value
for $R_{\omega\Lambda}$ and the arrows that for $R_{\phi\Lambda}$,
while the lines labeled DD2$-b$ (SFHo$-b$) were obtained with
$R_{\omega\Lambda}=1$.  The lower lines for each model correspond to
models containing the entire baryon octet, whereas the upper lines
correspond to models with $\Lambda$-hyperons only. In the sense that
the onset of hyperons softens the EoS, and the softening is stronger
the larger the number of hyperons included, the upper and lower curves
of each model limit the maximum mass obtained with any hyperonic model
almost independently of the details of the couplings of { $\Xi$-} and
$\Sigma$-hyperons which are even less known than that for $\Lambda$s
due to the lack of relevant experimental data. 

The lower curves of the
top panel of Fig.~\ref{fig:DD2hypernuclei} were obtained taking the
same couplings to the $\Lambda$-hyperon as model DD2$-a$ (SFHo$-a$), and for the
{ $\Xi$} and $\Sigma$ choosing the $SU(6)$-coupling to the
$\omega$-meson and fitting the coupling to the $\sigma$ -meson to
obtain a single particle potential in nuclear matter of $-18$ MeV
($-14$ MeV) and
$+30$ MeV, respectively.  The coupling of these hyperons to the
$\sigma^*$ and $\phi$ is taken equal to zero, and, therefore, we
consider that the maximum mass determined with this prescription will
define a lower limit, since, due to the dominance of the vector meson
at high densities, it is expected that its repulsive effect will
dominate over the attractive effect of $\sigma^*$.

It is obvious that taking $SU(6)$-values for the $\omega$-couplings
and not including the $\phi$-meson for { $\Xi$-} and $\Sigma$-hyperons
does not allow to reproduce the 2 $M_\odot$ constraint. 
{  For DD2, we have,
therefore, considered the effect of keeping the coupling of
the $\Lambda$ to all mesons as in DD2$-a$ and, besides,  coupling the $\Sigma$
and $\Xi$ also to the $\phi$-meson keeping the $SU(6)$-values
  for the vector mesons,  these models are labelled
  DD2$-c$ in Fig.~\ref{fig:DD2hypernuclei} top-left panel. Under these
  conditions, it is possible to describe
  two solar mass stars for a large enough $R_{\phi\Lambda}$. However, for SFHo, even taking only the
  $\Lambda$-hyperon with the $\Lambda$-meson coupling  calibrated with the $SU(6)$-values
  for the $\omega$ does not allow for a two solar mass star. Only
  breaking the SU(6) symmetry for both $\omega$ and $\phi$ vector-mesons the two solar
  masses constraint is satisfied.
}

In Table~\ref{tab:double} we give, for two values of
$R_{\omega\Lambda}$ and two values of $R_{\phi\Lambda}$ the calibrated
values of $R_{\sigma\Lambda}$ and $R_{\sigma^*\Lambda}$. For
reference, the corresponding values of the $\Lambda$-potential in
symmetric baryonic matter at saturation
$U_\Lambda^{(N)}(n_{\mathit{sat}})$ obtained from Eq.~(\ref{Ujk}) are
given as well as $U_{\Lambda}^{(\Lambda)}(n_{\mathit{sat}})$, and
$U_{\Lambda}^{(\Lambda)}(n_{\mathit{sat}}/5)$, the $\Lambda$
single-particle potential in $\Lambda$-matter.  It is interesting to
notice that for DD2 $U_{\Lambda}^{(\Lambda)} (n_{\mathit{sat}}/5) $is
close to the value of $-5$ MeV, as generally used in the literature to
fix these couplings.

In order to better understand the predictions obtained for the neutron
stars mass-radius relation, we display in
Fig.~\ref{fig:DD2hypernuclei} (bottom panels) the complete $M$-$R$
curves and illustrate with a hashed region the region limited by the
upper and lower limits shown in the top panels of the same figure. The
black line represents pure nucleonic stars. 
 We also include in the figure the
constraints imposed by the two pulsars PSR J1614-2230
{\citep{Demorest2010,Fonseca2016}} and  PSR J0348+0432
{\citep{Antoniadis2013}}. All configurations obtained with the larger
$\omega$ couplings are within the observed masses, while for the
$SU(6)$-couplings some configurations could be too small.

  { 
 The green line in the bottom-left panel labelled
DD2$-c$ was calculated taking for the  $\phi$-couplings the
  $SU(6)$-values. A maximum mass of $1.99\,M_\odot$ is obtained, just
slightly smaller than the maximum mass determined with DD2Y 
  and clearly above DD2Y$\sigma^*$, where { $\Xi$-} and
  $\Sigma$-hyperons couple to $\sigma^*$, too. This last difference
  can be attributed to the fact that DD2Y$\sigma^*$ has a much larger fraction
  of negatively charged hyperons and, therefore, smaller amount of
  electrons.

For the SFHo model, two extra $M-R$ curves are shown in the bottom-right panel with different choices for the
    vector-meson couplings, the models labelled SFHo$-c$
    (green) and SFHo$-d$ (cyan).  The  SFHo$-c$ model
    with calibrated $\Lambda$-couplings to hypernuclei and the
    $\Sigma$ and $\Xi$ potentials in symmetric matter as above,  has  ratios $R_{\omega i}=1$ for all
  hyperons and $R_{\phi\Lambda}=R_{\phi\Sigma}=-0.707$ and
  $R_{\phi\Xi}=-1.77$,  and predicts a maximum mass of $2.03 \,M_\odot$, slightly
above the one  given by SFHoY, very close to the maximum mass obtained
including only calibrated $\Lambda$s, see the top curves for SFHo$-b$ in
Fig.~\ref{fig:DD2hypernuclei} top right pannel. The SFHo$-d$ model was obtained with the calibrated
$\Lambda$ parameters with  $R_{\Lambda\omega}=1$ and $R_{\phi\Lambda}=-0.707$ and the
couplings to the $\Sigma$ and $\Xi$ as in the SFHoY model. Star
properties obtained with this model are very similar to the ones
obtained with SFHoY.}

\subsection{Hyperon content and thermodynamic properties}
{ Let us now discuss the properties of homogeneous matter
  obtained within the different EoS modesl, DD2Y, SFHoY and
  SFHoY$^*$.} Although there are small quantitative differences, the
regions in temperature and baryon number density where the overall
hyperon fraction exceeds $10^{-4}$, see Fig.~\ref{fig:contour}, have a
very similar shape in the different models.  The bump in the curves,
i.e., the part of the lines above approximately 20~MeV, where the
abundance of hyperons is still below $10^{-4}$, arises from the
competition between light nuclear clusters and hyperons in this
particular temperature and density domain and does not exist in the
EoSs built on nuclear models without light clusters,
see~\cite{OertelRMP16}. In a more complete model, where light clusters
and hyperons are allowed to coexist, the bump would probably disappear
and above a temperature of roughly 20 MeV, hyperons would exist at any
density.  {In \cite{Menezes2017}, hyperon fractions have been
  calculated in the presence of heavy clusters. Under these conditions
  these fractions were always below $10^{-5}$.  However, heavy
  clusters melt at quite low temperatures, so it is important to also
  include light clusters explicitly.}  At large temperatures results
are insentive to the interaction but do depend on the number of
competing species, such that the hyperon onset temperature is very
similar within all models.

The main differences between
models occur at low temperatures when the results are sensitive to the
hyperon-meson interactions. Thus, in particular for cold neutron stars,
the hyperon onset density is lower in DD2Y than in SFHoY.  There are
two reasons for that. First, the additional repulsion needed in SFHoY
to be consistent with the $2 M_\odot$ constraint suppresses hyperonic
degrees of freedom at high density and low temperatures. Therefore in
model SFHoY$^*$, with $SU(6)$ couplings and thus less repulsion than
SFHoY, hyperons appear at lower densities. Second, the smaller
symmetry energy in SFHo compared with DD2, see Fig.~\ref{fig:esym},
effectively disfavors hyperons with respect to nucleons.

Let us discuss these assertions now in more detail.  In
Fig.~\ref{fig:xifuncnb} the hyperon fractions are plotted as a
function of the baryonic density for different electron fractions and
$T=30$ MeV.  DD2Y clearly has the largest overall hyperon
fractions. Again, this can be due to the larger couplings to vector
mesons in the SFHoY model and to the smaller symmetry energy of SHFo
to DD2 for all densities. Comparing SFHoY and SFHoY$^*$ corroborates
these arguments: the latter has larger hyperon fractions due to less
repulsive couplings, but does not reach DD2Y due to the smaller
symmetry energy. In all three models, due to the high temperature, the hyperon
onset density is very similar and strongly correlated with the
disappearance of nuclear clusters in favor of homogeneous matter.
Within DD2Y, above the onset density the overall hyperon fraction
strongly increases with density. This is slightly less true for
SFHoY$^*$ and much less for SFHoY, where the additional repulsion for
hyperons more strongly affects the results at high densities.  

\begin{figure}[ht]
\includegraphics[width=0.85\linewidth]{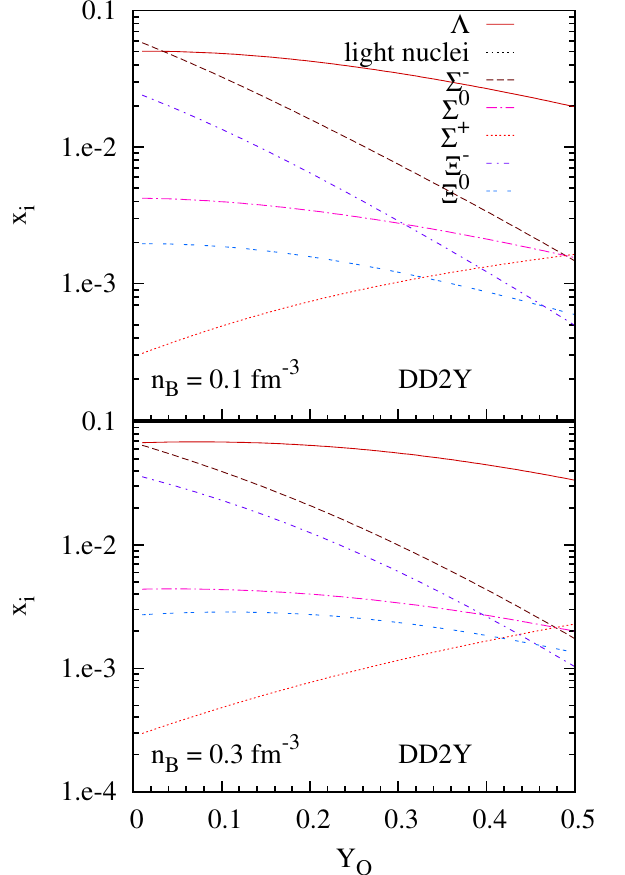}
\caption{(color online) 
  \label{fig:xifuncye} Hyperon fractions versus  charge fraction $Y_Q$
 for DD2Y for $T=50$ MeV and different baryonic densities.}
\end{figure}

\begin{figure*}[htb]
\includegraphics[width=1.\linewidth]{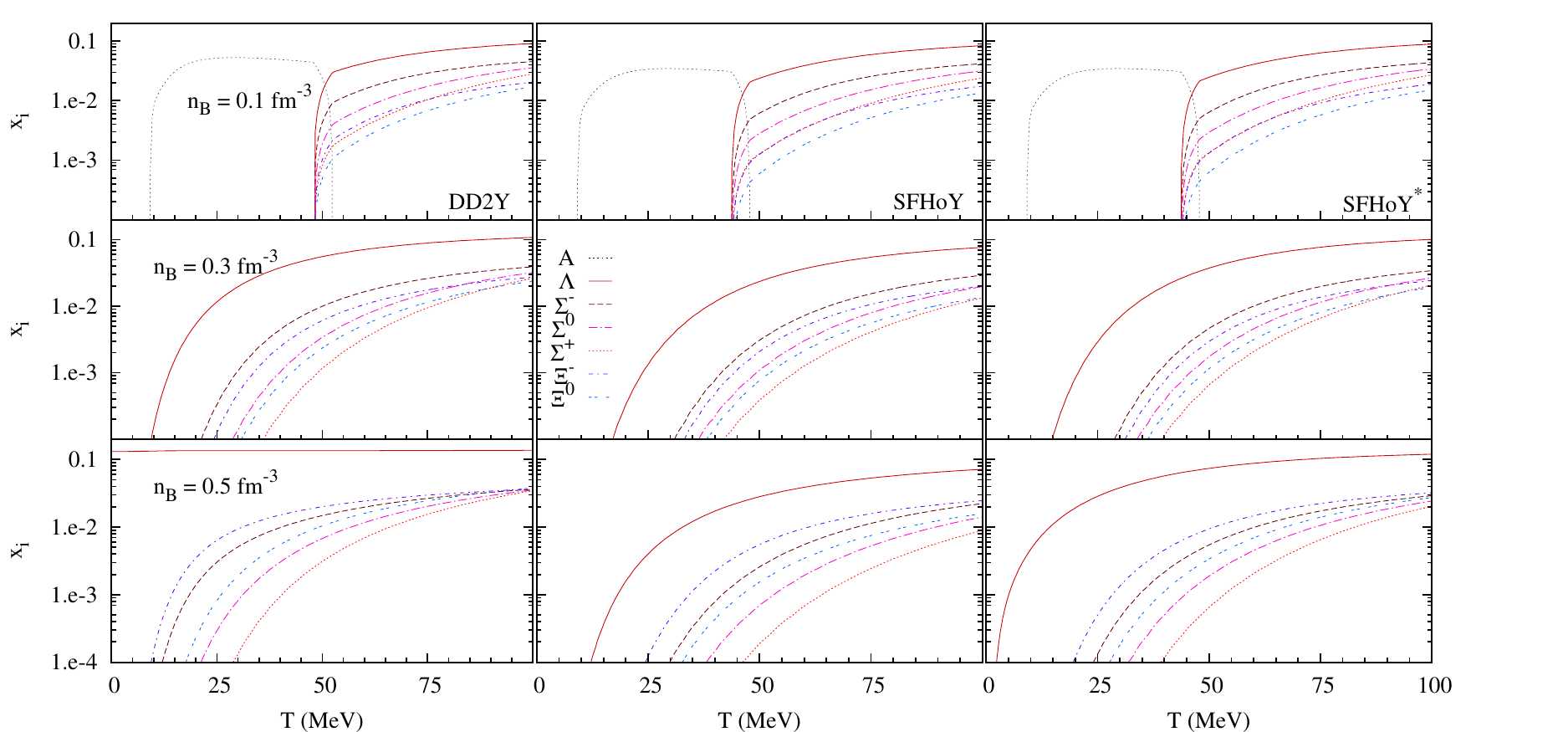}
\caption{(color online) Particle fractions as function of temperature
  for different values of fixed baryon number density and charge
  fraction $Y_Q = 0.3$ within DD2Y (left), SFHoY (middle) and SFHoY$^*$ (right) EoS. 
  \label{fig:xifunct}
}
\end{figure*}
The three most abundant hyperons obtained within all models at
$Y_Q=0.1$ coincide: $\Lambda$, $\Xi^-$ and $\Sigma^-$. $\Lambda$ has
the smallest mass and the abundance of the two other is due to the well
known negative isospin projection, favoring negatively charged
particles in matter with low charge fractions.  It is interesting to
see that $\Xi^-$ in SFHoY is as abundant as in DD2Y at large
densities, and contrary to DD2Y they become more abundant than
$\Lambda$-hyperons. The reason is that with our proposal the
couplings of the {$\Xi$} are less increased with respect to the
$SU(6)$ values than for the other hyperons, and as a consequence at
large densities and for very asymmetric matter the $\Xi$-hyperons become preferred even to the much
less massive $\Lambda$. In SFHoY$^*$, where $SU(6)$-couplings are
employed, still $\Lambda$-hyperons remain most abundant at high
densities, although the smaller symmetry energy in SFHo with respect
to DD2 more strongly favors negatively charged particles at low $Y_Q$.

With increasing charge fraction $Y_Q$, neutral and positively charged
hyperons and less massive ones are favored. For $Y_Q=0.5$, apart from
$\Lambda$-hyperons, the two most abundant hyperons are $\Xi^0$ and
$\Sigma^+$ followed by $\Xi^-$ for DD2Y. For SFHoY $\Xi^0$ and $\Xi^-$
are the most abundant and the $\Sigma$s are not present: again
repulsion is too strong for $\Sigma$-hyperons.

In Fig \ref{fig:xifuncye} we plot the hyperon fractions obtained
within DD2Y as a function of the charge fraction for $T=50$ MeV and
$n_B=0.1$ and 0.3 fm$^{-3}$. The hyperon $\Lambda$ is the most
abundant for all fractions shown except for very small fractions and
$n_B=0.1$fm$^{-3}$: this is the hyperon with the lowest mass and the
most bound in symmetric nuclear matter. However, the abundance of the
isoscalar $\Lambda$ only moderately depends on $Y_Q$, and for very
small $Y_Q$, $x_{\Sigma^-}$ exceeds $x_{\Lambda}$.  With decreasing
charge fraction, the charge chemical potential decreases, favoring
thus negatively charged particles, $\Sigma^-$ and $\Xi^-$. In the
present RMF models this is expressed via the coupling to the
$\rho$-meson. Inversely, $x_{\Sigma^+}$ increases with increasing
charge fraction.  On the other extreme, for $Y_Q\sim 0.5$, the
abundance of the different hyperon species mainly depends on their
mass, such that $x_{\Sigma^+}>x_{\Sigma^-}$. At this high
temperatures, the hyperonic interactions only marginally influence the
ordering of the hyperons.  At $T=0$, where interactions are important,
the first hyperon to set in is the $\Lambda$ followed by the
$\Xi^-$. The effect of temperature, which favors hyperons with smaller
masses, is larger for the smaller densities and this explains why the
difference between $\Sigma^-$ and $\Xi^-$ is larger for $n_B=0.1$
fm$^{-3}$ than for $n_B=0.3$ fm$^{-3}$. Results obtained within the
other two models are very similar, the ordering of the hyperons being
the same, the only difference being the abundances that are smaller,
in particular for SFHoY.

\begin{figure*}
\includegraphics[width=.8\linewidth]{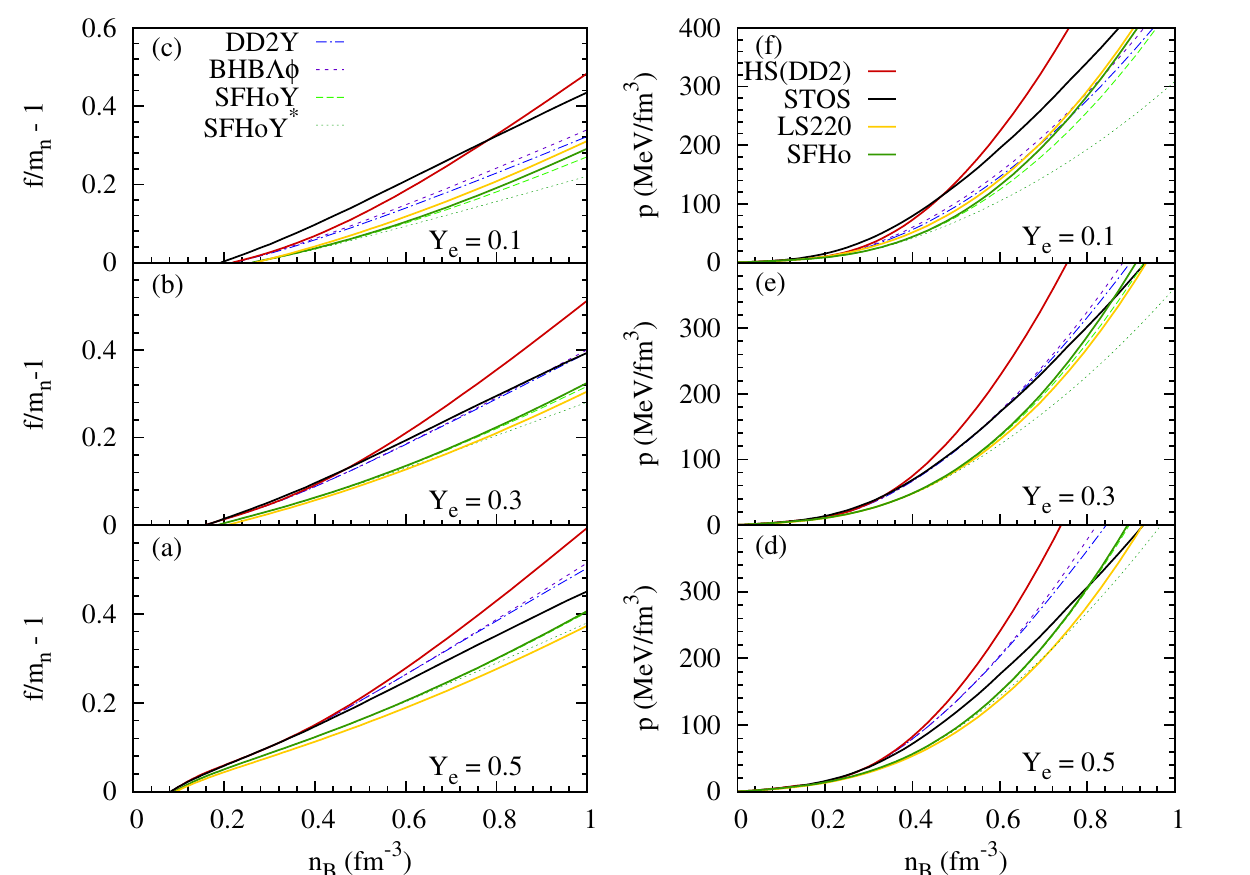}
\caption{(color online) Pressure (panels d-f) and normalised free
  energy per baryon (panels a-c) as function of baryon number density
  for different values of fixed electron fraction and $T = 30$ MeV
  within different EoS. Contributions from electrons/positrons and
  photons are included, demanding overall charge neutrality, i.e. $Y_e
  = Y_Q$.  For information, the pressure in the classical models LS220
  and STOS is displayed in addition.
  \label{fig:thermonb}
}
\end{figure*}
\begin{figure*}
\includegraphics[width=.8\linewidth]{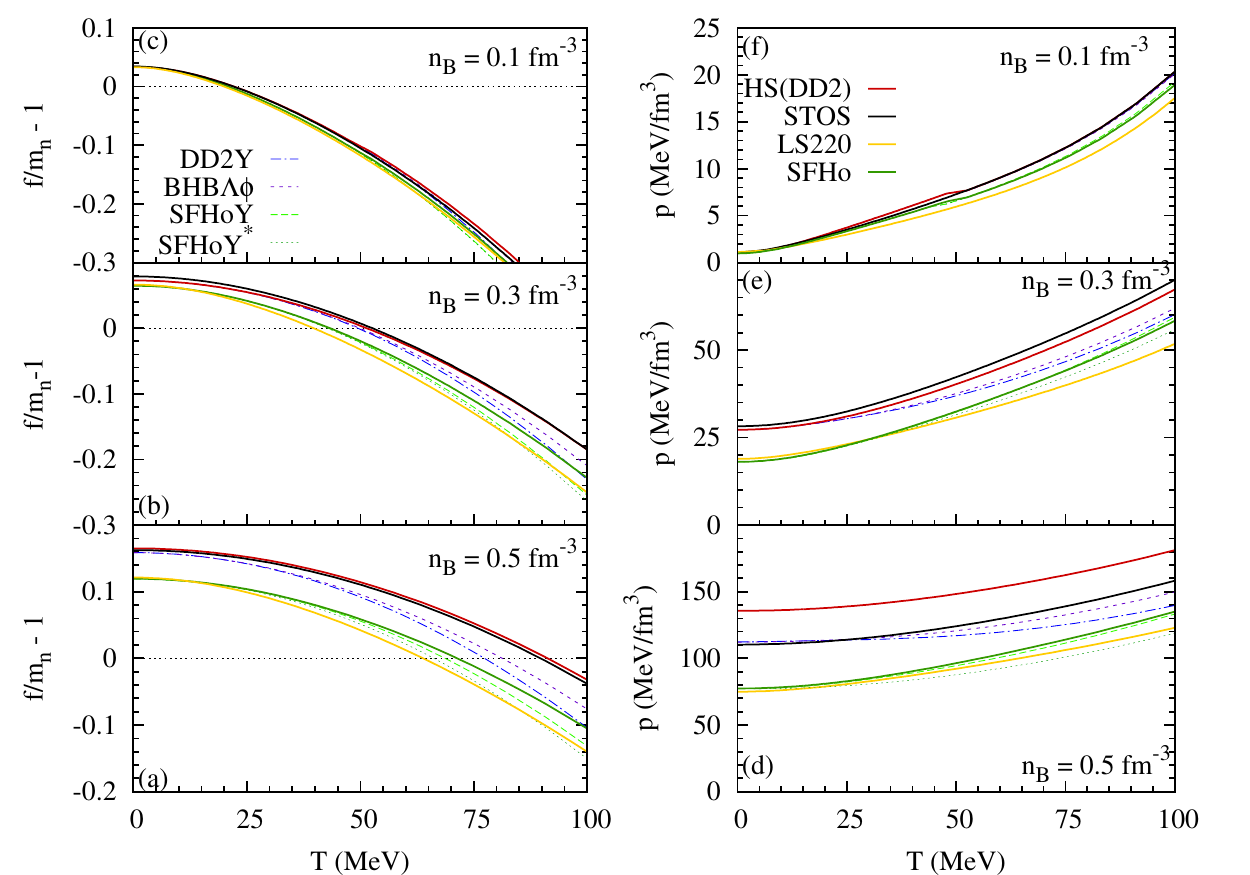}
\caption{(Color online) Same as Fig.~\ref{fig:thermonb}, but as function of temperature for $Y_e = 0.3$ and different fixed values of $n_B$.
  \label{fig:thermot}
}
\end{figure*}
The effect of temperature is better understood from
Fig. \ref{fig:xifunct}, where we display the particle fractions as a
function of temperature for a reference charge fraction, $Y_Q=0.3$,
and three values of the baryonic density. Some common features are
present in all the three models.  For the lowest temperatures, the
same sequence of hyperons occurs with respect to their abundance in
all models but the smallest abundance are obtained for SFHoY,
followed by SFHoY$^*$ and DD2Y. At large temperatues, the $\Sigma^+$
and the $\Xi^0$ become more abundant than their neutral or negatively
charged counterparts $\Sigma^0$ and $\Xi^-$ if the density is not too
large. This change of abundance will occur at lower temperatures for
DD2Y, followed by SFHoY$^*$ and finally SFHoY. A quite different
picture is described at $T=100$ MeV and $n_B=0.5$ fm$^{-3}$ by DD2Y
where all fractions, except for $\Lambda$ hyperons, coincide than by
the two other models. With SFHoY, the fraction of $\Sigma^+$ is still
the lowest and about 0.25 smaller than $x_{\Xi^-}$. This is a
consequence of the smaller symmetry energy in the two models based on
SFHo, and the larger repulsion felt by hyperons in SFHoY. In this
model, the same scenario of equal fractions is pushed to higher
temperatures and/or densities.  It is clear that the abundances
change as function of the different density values, and, in particular,
the most abundant hyperon after the $\Lambda$ is the $\Sigma^-$ for
the two lower densities and the $\Xi^-$ for $n_B=0.5$ fm$^{-3}$.  This
change is related to the fact that at low densities and high
temperatures, the interactions only slightly influence the
abundances, which are mostly given by masses. At high densities, as
seen here for $n_B = 0.5 \mathrm{fm}^{-3}$ (bottom panels), we
approach the situation at zero temperature discussed before where
interactions are important and $\Xi^-$ becomes the first hyperon to
set after $\Lambda$'s.

\begin{figure}
\includegraphics[width=1.\linewidth]{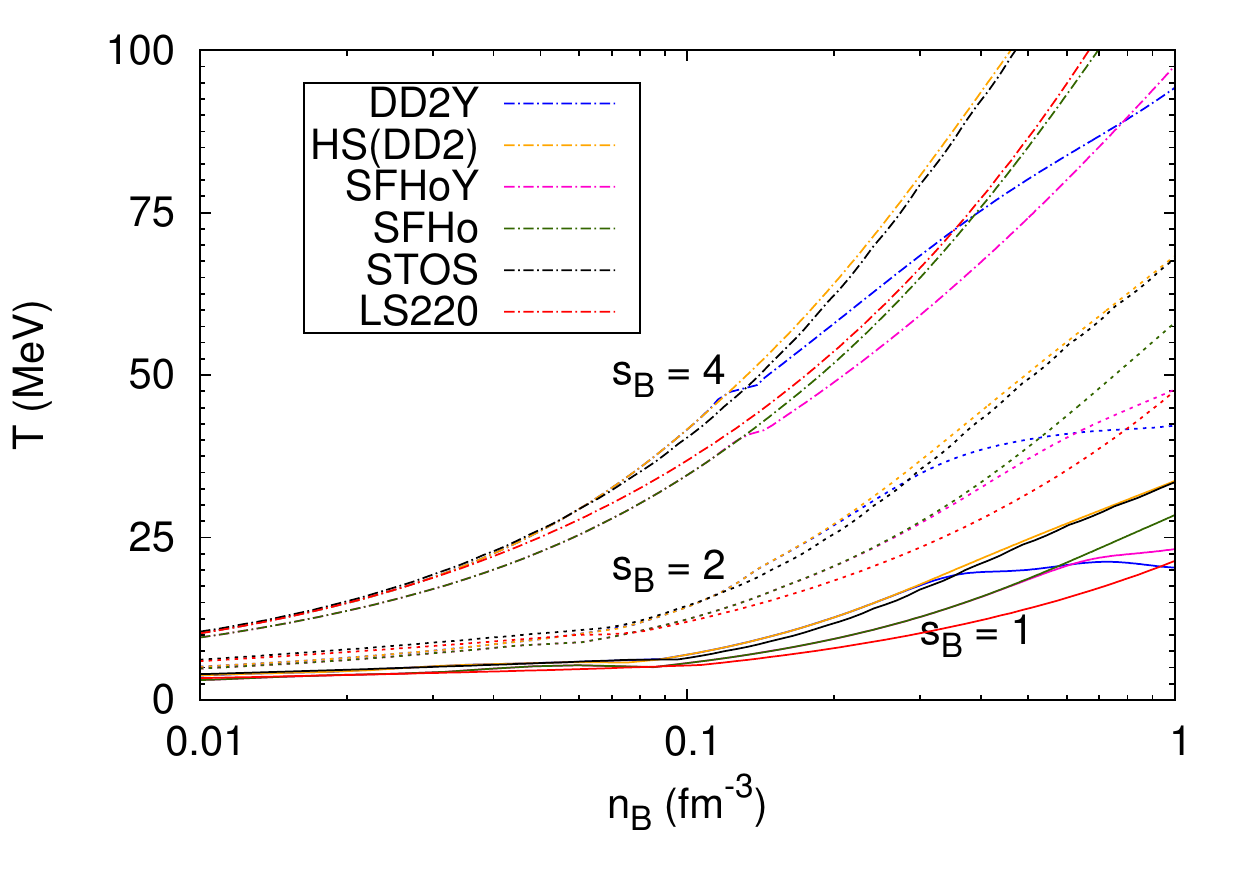}
\caption{(Color online) Temperature as function of baryon number
  density for different values of fixed entropy per baryon: $s_B=1$
  (plain lines), $s_B = 2$ (dashed lines) and $s_B = 4$ (dash-dotted
  lines) comparing purely nucleonic models with hyperonic ones,
  HD(DD2) and DD2Y as well as SFHo and SFHoY. The lepton fraction has
  been fixed to $Y_L=0.4$.
  \label{fig:tofn}
}
\end{figure}
\begin{figure}
\center\includegraphics[width=1.5\linewidth]{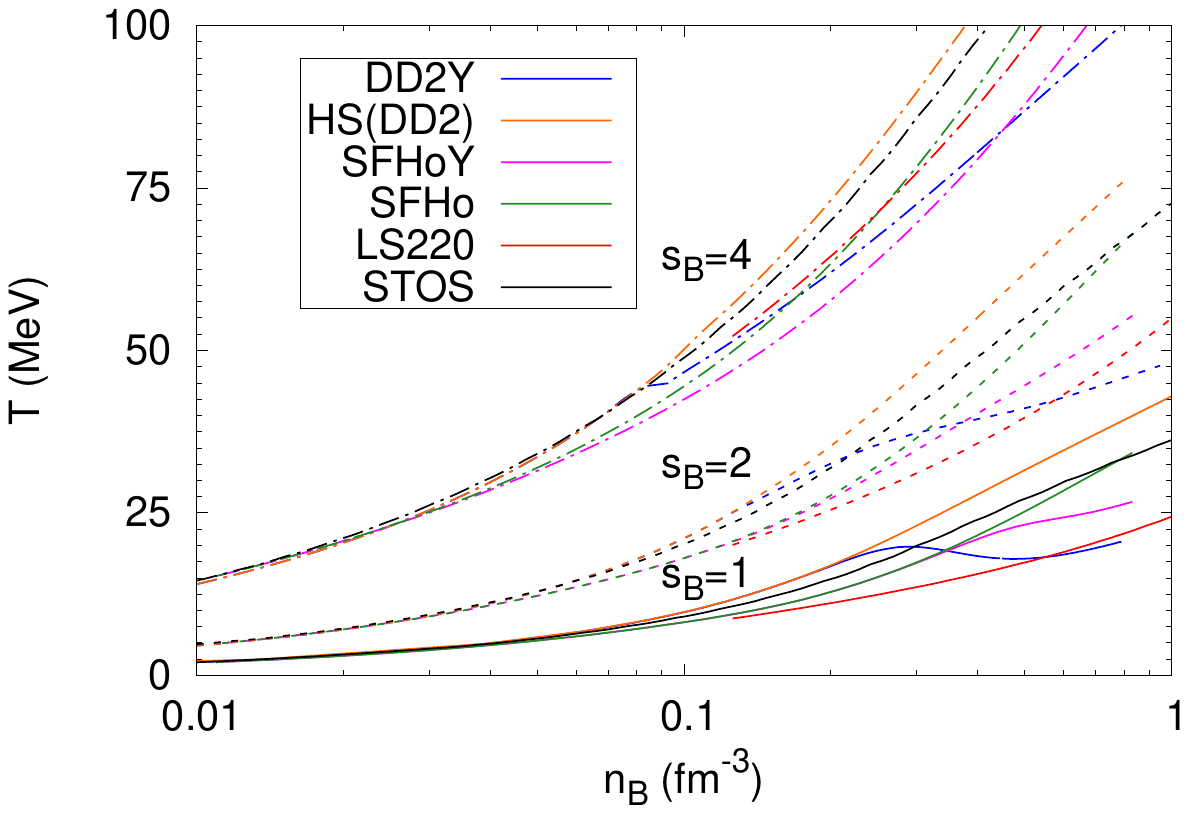}
\caption{(Color online) Same as Fig.~\ref{fig:tofn}, but considering neutrinoless $\beta$-equilibrium instead of $Y_L = 0.4$. 
  \label{fig:tofn2}
}
\end{figure}

Obviously, large hyperon abundances have strong effects on
thermodynamic quantities, too. 
Hence, as can be seen from Fig.~\ref{fig:thermonb}, pressure and free energy per baryon are considerably reduced above
roughly 2-3 times nuclear saturation density in the models with
hyperons compared with the purely nucleonic ones. 
The reduction is most important for low
electron fractions. This is clearly understandable since, as seen
before, the overall hyperon fractions are highest for small electron
fractions.  The higher hyperon fractions in DD2Y compared with
SFHoY$^*$ and in particular SFHoY explain the larger reduction within
DD2Y (and SFHoY$^*$), too. As already discussed in \cite{Marques2017},
only a small further reduction in DD2Y with respect to
BHB$\Lambda\phi$ can be observed. This is due to the fact that the
overall hyperon fraction is very similar in both models, the
additional hyperonic degrees of freedom in DD2Y being compensated by a
higher $\Lambda$-fraction in BHB$\Lambda\phi$. When comparing SFHoY
and SFHoY$^*$ the reduction is larger in the second model because
higher hyperon fractions are present, due to the less repulsive
hyperonic interactions.

In Fig.~\ref{fig:thermot}, pressure and free energy are plotted as
function of temperature for different values of $n_B$ and an
intermediate value of $Y_e = 0.3$. The results again confirm our above
findings: the presence of hyperonic degrees of freedom reduce pressure
and free energy with its impact increasing with temperature, as the
hyperon fractions increase. Again, the effect is less pronounced in
SFHoY since for the density values shown the more strongly repulsive
hyperonic interactions diminish the influence of hyperons within this
model. As can be seen comparing SFHoY$^*$ and DD2Y, the smaller
symmetry energy within SFHo also plays a non-negligible role in
rendering hyperons less important within the models based on SFHo.

The fraction of hyperons and the way they are distributed among the
different species at a given temperature will affect the entropy per
baryon, $s_B$, of the system, too. If the energy is shared among an
increased number of degrees of freedom, thermal excitations will be
reduced for each of them and entropy will be higher for systems with
hyperons than for purely nucleonic ones at a given temperature. On the
other hand, for a given $s_B$, we expect a lower temperature in
systems with hyperons. This is confirmed by the results shown in
Figs.~\ref{fig:tofn} and \ref{fig:tofn2}, where temperature as
function of baryon number density for three different values of
entropy per baryon ($s_B=1,\, 2,\, 4$) is displayed. These entropy
values correspond to typical values in proto-neutron stars. In
Fig.~\ref{fig:tofn}, the overall lepton fraction (including neutrinos)
has been fixed to $Y_L = 0.4$, again a typical value for proto-neutron
stars before neutrinos diffuse out of the star. In
Fig.~\ref{fig:tofn2}, the same quantities are plotted for
$\beta$-equilibrated stellar matter with no trapped neutrinos.  In
this figure we also include for reference the results for LS220 and
STOS. As expected, as soon as hyperons set in, the temperature
drops. This effect is more pronounced within DD2Y due to the larger
amount of hyperons. Comparing Figs.~\ref{fig:tofn} and
\ref{fig:tofn2}, it can be seen that for matter without neutrinos,
temperatures are generally larger since due to the absence of
neutrinos, there is one degree of freedom less. In this case hyperons
set in at very low densities, the larger the entropy the lower the
onset density.

It is interesting to notice that already if only nucleons are
considered, the two models show non negligible differences. SFHo
predicts lower temperatures than DD2 both for matter with trapped and
neutrino free matter (see Figs.~\ref{fig:tofn} and
\ref{fig:tofn2}). This is probably due to the differences in the
symmetry energy: a smaller symmetry energy leads to lower proton
fractions and, thus, electron fractions, leading to a lower
temperature for a given entropy per baryon. The effect is slighlty
more pronounced at fixed lepton fraction, since a lower electron
fraction corresponds to a larger neutrino fraction within the SFHo
model, too, lowering additionally the temperature needed for a given
entropy per baryon. Comparing STOS and LS220 results with the above
ones we observe that: 
a) the temperatures predicted by STOS are similar to DD2, being only slightly smaller at large
densities; b) LS220 predicts smaller temperatures than the other nuclear models. The temperatures remain even below those predicted by the hyperonic models in a large density range.  

\section{SUMMARY AND CONCLUSIONS}

Two new general purpose EoS applicable within neutron star merger and
core-collapse simulations and including hyperons were proposed and
discussed. The two EoS are based on the nucleonic EoS
SFHo~\citep{steiner13}. The entire baryonic octet is considered and
the hyperonic interaction is described in the standard way for RMF
models, i.e. mediated by $\sigma$, $\omega$ and $\rho$-mesons plus the
isoscalar-vector meson with hidden strangeness $\phi$. In one of the
models the couplings of the isoscalar-vector mesons were obtained
imposing $SU(6)$- symmetry, SFHoY$^*$, as done in most models
including hyperons due to the lack of experimental information on
hyperonic couplings. Since within this model the maximum mass of a
cold $\beta$-equilibrated neutron star is clearly below $2 M_\odot$,
in our second model, SFHoY, the $SU(6)$-constraint has been relaxed
and more repulsive couplings were chosen. Within the latter
parametrization, the maximum mass is compatible with the constraints
imposed by the two pulsars PSR
J1614-2230~\citep{Demorest2010,Fonseca2016} and PSR
J0348+0432~\citep{Antoniadis2013}. To determine the isoscalar scalar
hyperonic couplings, we have considered the following values for the
hyperonic single particle potentials in symmetric nuclear matter:
$U_{\Lambda}^{(N)}(n_{\mathit{sat}}) = $-30 MeV,
$U_{\Sigma}^{(N)}(n_{\mathit{sat}}) = $+30 MeV,
$U_{\Xi}^{(N)}(n_{\mathit{sat}}) = $-14/-18 MeV. 
We have shown, 
following the lines of \cite{Fortin2017} that the chosen
interactions are compatible with properties of
single $\Lambda$-hyperons. For future extension of our work we have
indicated other  { parameterisations based on both the DD2
  and SFHo models, which in addition allow to reproduce
the bond energy of $_{\Lambda\Lambda}^6$ He and to describe for two solar
mass neutron stars}. Our new EoS are available
in tabular form from the \textsc{Compose} database \citep{Typel2013},
see appendix~\ref{app:tables}.

The DD2Y EoS proposed in \cite{Marques2017} and our new model SFHoY
are the only general purpose EoS models containing the entire baryon
octet up to now well compatible with the relevant constraints on the
EoS. One of the main differences between these two models is the much
softer symmetry energy in the underlying nuclear model SFHo than in
DD2. Together with the additional repulsion needed in SFHoY to obtain
a $2 M_\odot$ cold neutron star this leads to much lower overall
hyperon fractions in SFHoY than in DD2Y. Consequently the effects on
thermodynamic properties are much less pronounced in SFHoY.

In view of all above results, we may expect a different proto-neutron
star evolution and an impact on neutron star merger dynamics from
including hyperonic degrees of freedom within the EoS. Indications can
be found from the simulations of black hole (BH) formation showing a
reduced time until collapse to a BH (see e.g. \cite{Peres_13}) in the presence
of hyperons, but further studies are in order with EoS models, such as
those presented here, allowing for all hyperons and being compatible
with constraints, in particular a 2$ M_\odot$ cold neutron star.

\begin{acknowledgements}
This work has been partially funded by the ``Gravitation et physique
fondamentale'' action of the Observatoire de Paris, by Funda\c c\~ao
para a Ci\^encia e Tecnologia (FCT), Portugal, under the project
No. UID/FIS/04564/2016, the Polish National Science Centre (NCN) under
grant No. UMO-2014/13/B/ST9/02621, and the COST action MP1304
``NewComsptar''.
\end{acknowledgements}

\bibliographystyle{pasa-mnras}
\bibliography{biblio}
\begin{appendix}
\section{Combining different parts of the EoS}
The HS(DD2) and the SFHo EOS contain the transition from inhomogeneous or
clusterized matter to uniform nucleonic matter. This is done via the
excluded volume mechanism, which suppresses nuclei around and above
nuclear saturation density. On top of that, for some thermodynamic
conditions a Maxwell construction over a small range in density is
necessary, for details see~\cite{hempel12}.

Here the situation is slightly more complicated, since homogeneous
matter might contain hyperons. In the simplest case, hyperons appear
within homogeneous (nucleonic) matter and it is sufficient to minimize
the free energy of the homogeneous system to decide upon the particle
content of matter. Such a situation occurs at low temperatures and
high densities.

In some parts of the $T$-$n_B$ diagram, however, a transition from
inhomogeneous matter directly to hyperonic homogeneous matter is
observed. This is the case at low densities and high temperatures,
i.e. the density regions up to the bumps in
Fig.~\ref{fig:contour}. There, light clusters compete with hyperonic
degrees of freedom with only very small differences in free energy
which are of the order of the numerical accuracy of the EoS
calculation. To technically construct the transition in this region,
we follow a similar prescription as in~\cite{Banik:2014qja, Marques2017} and
introduce a threshold value for the total hyperon fraction,
$Y_{\mathit{hyperons}} = \sum_{j \in B_Y} n_j/n_B$.  We let hyperonic
matter appear only if $Y_{\mathit{hyperons}} > 10^{-6}$. Note that the
hyperon fraction is not the same as the strangeness fraction, $Y_S$,
defined as the sum of all particle fractions multiplied by their
respective strangeness quantum numbers, $Y_S = \sum_{j \in B} S_j
n_j/n_B$.

Although the above described procedure allows to construct a smooth
transition between the different parts of the EoS, it is of course not
completely consistent. In principle, whenever hyperons compete with
light nuclear clusters, the free energy of the system should be
minimized allowing simultaneously for all different possibilities,
e.g. a coexistence of light clusters with hyperons. In view of the
tiny differences in free energy and the small fractions of particles
other than nucleons, electrons, and photons in the transition region,
a completely consistent treatment is left for future work.

\section{Technical issues of the EoS tables}
\label{app:tables}
The EoSs DD2Y and SFHoY are  provided in a tabular form in the
\textsc{Compose} data base, \url{http://compose.obspm.fr} as a
function of $T, n_B, Y_e$, either including the contribution from electrons and photons or containing only the baryonic part. Note that the \textsc{Compose} software allows to calculate additional quantities such as, e.g., sound speed, from those provided in the tables. Please see the \textsc{Compose} manual~\citep{Typel2013} and
the data sheet on the web site for more details about the definition
of the different quantities.

\begin{itemize}
\item The grid is specified as follows (with the first value in the density grid for DD2Y and the second for SFHoY and SFHoY$^*$):
\begin{table}[h!]
\begin{tabular}{c||c|c|c}
 & $T$ & $n_B$ & $Y_e$ \\ \hline
\# of points & 80 & 302/308 & 59 \\
Minimum & 0.1 MeV & $10^{-12} \mathrm{fm}^{-3}$ & 0.01\\
Maximum & 158.5 MeV & $1.202/1.9~\mathrm{fm}^{-3}$ & 0.6\\
Scaling & logarithmic & logarithmic & linear
\end{tabular}
\end{table}
\end{itemize}
\begin{itemize}
\item Thermodynamic quantities provided:
\begin{enumerate}
\item Pressure divided by baryon number density $p/n_B$ [MeV]
\item Entropy per baryon $s/n_B$
\item Scaled baryon chemical potential $\mu_B/m_n - 1$
\item Scaled charge chemical potential $\mu_Q/m_n$
\item Scaled (electron) lepton chemical potential $\mu_L/m_n$
\item Scaled free energy per baryon $f/(n_B m_n) - 1$
\item Scaled energy per baryon $e/(n_B m_n) - 1$
\end{enumerate}
\item Compositional data provided:
\begin{enumerate}
\item Particle fractions of baryons and electrons, $Y_i = n_i/n_B$
\item Particle fractions of deuterons ($^2$H), tritons ($ ^3$H), $^3$He, and $\alpha$-particles ($ ^4$He)
\item Fraction of a representative (average) heavy nucleus, together with its average mass number and average charge
\end{enumerate}
Please note that only nonzero particle fractions are listed.
\item Effective Dirac masses $M^*$ of all baryons with nonzero density
  are provided within homogeneous matter.
\end{itemize}
\end{appendix}
\end{document}